%% file: main.tex
\documentclass[aps, prc, nofootinbib, superscriptaddress]{revtex4-1}
\input{preamble}

\graphicspath{{figures/}}

\begin{document}

\input{FrontMatter/FrontMatter_main}

\input{introduction}

\input{background/background_main}

\input{lagrangian}

\input{calculations/calculations_main}

\input{conclusion}

\input{acknowledgments}

\input{Appendix/appendix_main}

\bibliography{references}
\end{document}

%% file: preamble.tex
\usepackage[utf8]{inputenc}
\usepackage{graphicx}
\usepackage[dvipsnames]{xcolor}
\usepackage[margin=0.7in]{geometry}
\usepackage{float}
\usepackage{bm}
\usepackage{amsmath, amsthm, amssymb, amsfonts}
\usepackage{bbm}
\usepackage{slashed}
\usepackage{physics}
\usepackage{epsfig}
\usepackage{xspace}
\usepackage{graphics}
\usepackage{slashed}
\usepackage{bbold}
\usepackage{bbm}
\usepackage{empheq}
\usepackage{mathrsfs}
\usepackage{xcolor}
\usepackage{cancel}
\usepackage{comment}
\usepackage[colorlinks=true, citecolor = blue, linkcolor = cyan]{hyperref}
\newcommand{\Nc}{\ensuremath{N_c}\xspace}
\newcommand{\LONc}{\ensuremath{\text{LO-in-\Nc}}\xspace}

\newcommand{\oneS}{{{}^{1}\!S_0}}
\newcommand{\threeS}{{{}^{3}\!S_1}}

\newcommand{\Galilean}{\overset{\leftrightarrow}{\nabla}}

\newcommand{\del}{\nabla}

\newcommand{\calO}{\ensuremath{\mathcal{O}}}

\newcommand{\calI}{\ensuremath{\mathcal{I}}}
\newcommand{\calG}{\ensuremath{\mathcal{G}}}

\newcommand{\calL}{\ensuremath{\mathcal{L}}}
\newcommand{\calM}{\ensuremath{\mathcal{M}}}

\newcommand{\1}{\mathbbm{1}}

\newcommand{\nopi}{\ensuremath{\pi\hskip-0.40em /}}
\newcommand{\eftnopi}{EFT$_{\nopi}$\xspace}
\newcommand{\LambdaNoPion}{\Lambda_{\nopi}\xspace}

\newcommand{\Ndag}{N^\dagger}

\newcommand{\beq}{\begin{equation}}
\newcommand{\eeq}{\end{equation}}

\newcommand{\lb}{\left(}
\newcommand{\rb}{\right)}

\newcommand{\NN}{\ensuremath{N\!N}\xspace}

\usepackage{ulem}
\newcommand\redsout{\bgroup\markoverwith{\textcolor{red}{\rule[0.5ex]{2pt}{1.4pt}}}\ULon}

\renewcommand{\emph}[1]{\textit{#1}}

%% file: FrontMatter/FrontMatter_main.tex
\input{FrontMatter/info}

\preprint{MITP-22-089}

\input{FrontMatter/abstract}

\maketitle

%% file: FrontMatter/info.tex
\title{Large-\Nc constraints for elastic dark matter-light nucleus scattering in pionless effective field theory}
\author{Thomas R.~Richardson}
\email{trr20@duke.edu}
\affiliation{Department of Physics and Astronomy, University of South Carolina, Columbia, South Carolina 29208, USA}
\affiliation{Department of Physics, Duke University, Durham, North Carolina 27708, USA}

\author{Xincheng Lin}
\email{xincheng.lin@duke.edu}
\affiliation{Department of Physics, Duke University, Durham, North Carolina 27708, USA}
\author{Son T. Nguyen}
\email{stn7@phy.duke.edu}
\affiliation{Department of Physics, Duke University, Durham, North Carolina 27708, USA}
\date{\today}

%% file: FrontMatter/abstract.tex
\begin{abstract}
    Recent proposals for the use of light nuclei as dark matter direct detection targets necessitate a strong theoretical understanding of the nuclear physics involved.
    We perform relevant calculations for dark matter-light nucleus scattering in a combined pionless effective field theory and  large-\Nc expansion, where \Nc is the number of quark colors.
    We include a general set of one-nucleon currents that have been used in other effective theories, as well as novel two-nucleon contact currents.
    First, we obtain constraints for the relative sizes of the dark matter couplings to the one- and two-nucleon currents through the large-\Nc expansion.
    Then, we use these constraints to make predictions for the relative sizes of spin-dependent and spin-independent cross sections for dark matter scattering off of a nucleon, a deuteron, a triton, and helium-3.
\end{abstract}

%% file: introduction.tex
\section{Introduction}
    \label{sec:Introduction}

Although evidence for the existence of dark matter is abundant (see, e.g., \cite{Feng:2010gw, Roszkowski:2017nbc, Schumann:2019eaa, ParticleDataGroup:2020ssz} for review), the detection of dark matter is still a high priority in experimental searches for Beyond the Standard Model (BSM) physics. 
There are several complementary searches dark matter interacting with nuclei including underground direct detection experiments \cite{PandaX_2017,LUX_2016,Deap_2018,Xenon1T_2017,CDMS_2018, XENON:2019gfn, aprile_spin-dependent_2019}, cosmology \cite{boddy_first_2018, gluscevic_constraints_2018, maamari_bounds_2021}, and spherical proportional counters \cite{guo_concept_2013, ito_scintillation_2013, gerbier_NEWS_2014, profumo_GeV_2016, hertel_direct_2019,Maris_Helium_Evaporation_2017}.
Large direct detection experiments attempt to measure nuclear recoil due to dark matter scattering off of a range of targets from fluorine to xenon, spherical detectors make use of light nuclei such as helium, while cosmological probes can directly constrain the dark matter-proton interaction.
These avenues can be used to detect weakly interacting massive particles (WIMPs), which are dark matter candidates with mass $m_\chi \gtrsim O(1)$ GeV \cite{Feng:2010gw, Roszkowski:2017nbc, Schumann:2019eaa, ParticleDataGroup:2020ssz} while cosmological and spherical detection searches are also sensitive to low mass dark matter.
The potential to use light nuclei as detection targets can also be realized through doping xenon with hydrogen, deuterium, or helium as suggested in Ref.~\cite{Aalbers:2022dzr}.

Given the recent development of detectors that use light nuclei as targets and the maturing constraints from cosmology, it is important to obtain a clear theoretical picture of few-nucleon physics in the presence of external WIMP fields.
Effective field theory (EFT) provides a model-independent framework to achieve this understanding. 
Several studies in this vein have been conducted in chiral EFT (ChEFT) \cite{prezeau_new_2003, cirigliano_wimp-nucleus_2012, menendez_spin-dependent_2012, Klos_large-scale_2013,hoferichter_chiral_2015, korber_first-principle_2017, hoferichter_nuclear_2019, bishara_chiral_2017, bishara_from-quarks_2017, hoferichter_analysis_2016, Andreoli:2018etf}, a nonrelativistic EFT for the nucleus as a whole \cite{fan_non-relativistic_2010}, and a nonrelativistic EFT for single nucleons \cite{fitzpatrick_effective_2013, fitzpatrick_model-independent_2012, anand_model-independent_2015, anand_weakly_2014, hill_standard_2015}.
For elastic dark matter-light nucleus scattering, the momentum transfer has an upper bound of a few MeV \cite{korber_first-principle_2017}, which is much less than the pion mass. 
Therefore, pionless EFT (\eftnopi), an EFT in which pions are integrated out, is well suited for these systems (see Refs.~\cite{vanKolck:1999mw, beane_hadrons_2001, bedaque_effective_2002, Platter:2009gz, Hammer:2019poc} for reviews).
In this work, we consider the nucleon, the deuteron, the triton (${}^3$H), and helium-3 (${}^3$He) as targets for dark matter detection in \eftnopi.

The use of \eftnopi departs from existing calculations in a few points. 
First, the renormalization and power counting of \eftnopi is well understood \cite{Yang:2019hkn}. Second, calculations of the properties of the deuteron can be performed analytically (see, e.g., Refs.~\cite{Kaplan_perturbative_1998,chen_nucleon-nucleon_1999,Chen_npdgamma_1999,Rupak_bigbag_2000}).
Calculations involving ${}^3$H or ${}^3$He, on the other hand, are still performed numerically as in ChEFT.
Third, while two-nucleon currents that arise due to pion exchange between the WIMP and the nucleons have been studied in ChEFT \cite{prezeau_new_2003, korber_first-principle_2017, cirigliano_wimp-nucleus_2012}, two-nucleon contact currents have been neglected because they are thought to be higher order in the power counting.
These contact currents have also been neglected in the nonrelativistic EFT of Refs.~\cite{fitzpatrick_effective_2013, fitzpatrick_model-independent_2012, anand_model-independent_2015, anand_weakly_2014, hill_standard_2015}.
However, when these currents connect two S-wave states, they receive an infrared enhancement in \eftnopi \cite{bedaque_effective_2002} and can yield significant contributions.
The possibility that the contact currents are enhanced in ChEFT as well was discussed in Ref.~\cite{hoferichter_nuclear_2019}.
In particular, there are examples of electroweak contact currents that contribute at next-to-leading order (NLO) in \eftnopi \cite{Kaplan_perturbative_1998, chen_nucleon-nucleon_1999,Chen_npdgamma_1999, butler_elastic_2000, butler_neutrino_2001, butler_proton-proton_2001}.
Therefore, it is necessary to include these currents from the outset in a complete NLO calculation.

In \eftnopi, like all EFTs, the Lagrangian is organized according to a power counting scheme. 
The Lagrangian possesses a finite number of operators at each order in the power counting; however, every operator in the effective Lagrangian is accompanied by an undetermined low energy coefficient (LEC).
These LECs, which are typically assumed to be of natural size once any relevant dimensional factors have been removed, must be determined from data or from nonperturbative quantum chromodynamics (QCD) calculations such as lattice QCD.
In the context of dark matter interactions with few-nucleon systems, the determination of the LECs is not yet possible.
Therefore, theoretical constraints from other sources are vital to guide the interpretation of available data and to prioritize lattice QCD calculations.

In this paper, we use the spin-flavor symmetry of baryons \cite{dashen_1/nc_1994, dashen_baryon_1993, dashen_spin_1995, carone_spin_1994, gervais_large-_1984, gervais_large-_1984-1, luty_baryons_1994} (see Ref.~\cite{Jenkins:1998wy} for review) in the large-\Nc limit of QCD \cite{t_hooft_planar_1974, t_hooft_two-dimensional_1974}, where \Nc is the number of quark colors, to constrain the relative sizes of the LECs at each order in the EFT power counting.
These constraints impose an expansion in powers of 1/\Nc on top of the EFT expansion.
Thus, the dual expansion reduces the number of LECs required in an EFT calculation to a given order in the combined power counting.
Specifically, we perform a large-\Nc analysis for both single-nucleon and two-nucleon currents relevant for WIMP-nucleus scattering.
While we consider \eftnopi explicitly, the contact currents in this work also appear in ChEFT so the large-\Nc constraints presented here are equally applicable.
Similar analyses have been performed for two-derivative two-nucleon contact terms \cite{schindler_large-$n_c$_2018}, parity violating two-nucleon contact terms \cite{schindler_large-$n_c$_2016, nguyen_large-N-three-derivative_2021}, time-reversal-invariance violation \cite{samart_time-reversal-invariance-violating_2016, vanasse_time-reversal-invariance_2019}, the three-nucleon potential \cite{phillips_three-nucleon_2013}, and two-nucleon currents in external magnetic and axial fields as well as lepton number and isospin violating currents \cite{richardson_large-_2020, richardson_large_2021}.

The paper is organized as follows: in Sec.~\ref{sec:pionless}, we discuss the key features of \eftnopi and the large-\Nc expansion. 
In Sec.~\ref{sec:large_Nc_analysis}, we introduce the effective Lagrangian with a WIMP coupled to one- and two-nucleon currents and discuss the large-\Nc scaling of the LECs. 
In Sec.~\ref{sec:elastic_scattering}, we calculate the parity ($P$) and time-reversal-invariance ($T$) conserving differential cross sections for a WIMP scattering off of a nucleon, a deuteron, ${}^3$H, and ${}^3$He. 
Then we explore the impact of the large-\Nc constraints for the LECs on the relative sizes of the cross sections and compare a selection of our results for a specific model to those of Refs.~\cite{korber_first-principle_2017, Andreoli:2018etf}. 
Conclusions are provided in Sec.~\ref{sec:conclusions}.
Appendix \ref{appendix:triton} contains details related to the dark matter-${}^3$H and ${}^3$He elastic scattering amplitudes, appendix \ref{appendix:dleft} presents the matching of quark currents to one-nucleon currents, and appendix \ref{appendix:dibaryon} presents the matching of dibaryon currents to two-nucleon currents.

%% file: background/background_main.tex
\input{background/EFT}

\input{background/LargeN}

%% file: background/EFT.tex
\section{The combined large-\Nc and pionless effective field theory expansion}
    \label{sec:pionless}
    
The \eftnopi Lagrangian consists of a tower of contact operators containing nucleon fields and external fields.
Each operator has an \textit{a priori} undetermined LEC that encapsulates physics above the breakdown scale $\LambdaNoPion \sim m_\pi$ \cite{kaplan_nucleon_1996, kaplan_new-expansion_1998, Kaplan:1998we, vanKolck:1998bw, chen_nucleon-nucleon_1999}.
If the characteristic momentum or momentum transfer of the system is $Q \ll m_\pi$, then the LECs will scale with a combination of powers of $Q$ and $\LambdaNoPion$ such that observables are expanded in powers of $Q/\LambdaNoPion$.

Because the scales of few-nucleon systems are fine tuned, power counting is more subtle than naive dimensional analysis \cite{kaplan_nucleon_1996, kaplan_new-expansion_1998, Kaplan:1998we, vanKolck:1998bw}.
In the two-nucleon system, we use dimensional regularization with the power divergent subtraction (PDS) scheme \cite{kaplan_new-expansion_1998}, which introduces an explicit dependence on the subtraction point $\mu$ taken to be on the order of $Q$ in the LECs. 
This scheme yields a consistent power counting that accounts for the fine-tuning of the $\NN$ interaction.
Furthermore, the LECs of the $\NN$  system may be fixed to reproduce the effective range expansion (ERE) about zero momentum \cite{kaplan_new-expansion_1998} or about the deuteron pole \cite{chen_nucleon-nucleon_1999}, or they may be fit to reproduce the residue of the $\NN$ amplitude about the deuteron pole exactly at NLO \cite{Phillips:1999hh, griesshammer_improved_2004}.
In the three-body system, amplitudes are obtained through numerical solutions of integral equations, which are regulated through a hard cutoff \cite{bedaque_effective_1998, bedaque_nucleon-deuteron_1998,bedaque_renormalization_1999,bedaque_three-boson_1999,bedaque_effective_2000,Hammer_renormalized_2000}.
Through these calculations, it was shown that a three-body contact term is required at LO for renormalization in processes in the doublet S-wave channel \cite{bedaque_renormalization_1999,bedaque_three-boson_1999,bedaque_effective_2000}.

%% file: background/LargeN.tex
The large-\Nc expansion can be combined with the \eftnopi expansion in order to further constrain the LECs at each order in the combined power counting.
Deriving large-\Nc constraints for nucleon matrix elements of QCD operators relies on the SU(4) spin-flavor symmetry of the baryon sector of large-\Nc QCD \cite{dashen_1/nc_1994, dashen_baryon_1993, dashen_spin_1995, carone_spin_1994, luty_baryons_1994}.
For single-baryon matrix elements, an $m$-quark operator can be expanded as \cite{dashen_spin_1995}
    \begin{equation}
            \label{eq:largeN:operator_expansion}
        \bra{B'} \calO^{(m)}_{\text{QCD}} \ket{B} = \bra{B'} \Nc^m \sum_{n, s, t} c^{(n)} \left( \frac{S^i}{\Nc} \right)^s \left( \frac{I^a}{\Nc} \right)^t \left( \frac{G^{ia}}{\Nc} \right)^{n - s - t} \ket{B},
    \end{equation}
where the one-body operators are
    \begin{equation}
            \label{eq:largeN:one-body_operators}
        S^i = q^\dagger \frac{\sigma^i}{2} q, \ I^a = q^\dagger \frac{\tau^a}{2} q, \ G^{ia} = q^\dagger \frac{\sigma^i}{2} \frac{\tau^a}{2} q \,  .
    \end{equation}
Since quarks in a nucleon are antisymmetrized in their color indices, $q$ is a colorless, bosonic quark field. 
The large-\Nc scaling for a general $n$-body operator with spin $S$ and isospin $I$ is \cite{kaplan_spin_1996, kaplan_nucleon-nucleon_1997}
    \begin{equation}
            \label{eq:largeN:n-body_scaling}
        \bra{B'} \frac{ \calO^{(n)}_{IS} }{ \Nc^n } \ket{B} \lesssim \Nc^{- \abs{I - S} } \ .
    \end{equation}
Therefore, the one-body operators scale as
    \begin{equation}
            \label{eq:largeN:none-body_scaling}
        \begin{split}
            \bra{B'} \frac{ \1 }{ \Nc } \ket{B} & \sim \bra{B'} \frac{ G^{ia} }{ \Nc } \ket{B} \lesssim 1 \ , \\
            \bra{B'} \frac{ S^i }{ \Nc } \ket{B} & \sim \bra{B'} \frac{ I^a }{ \Nc } \ket{B} \lesssim \frac{1}{\Nc} \ .
        \end{split}
    \end{equation}

In the large-\Nc limit, the two-nucleon interaction is $O(\Nc)$ and is described by a Hartree Hamiltonian \cite{witten_baryons_1979}, which has the expansion \cite{dashen_spin_1995, kaplan_spin_1996, kaplan_nucleon-nucleon_1997}
    \begin{equation}
            \label{eq:largeN:Hartree}
        H = \Nc \sum_{n,s,t}  v_{stn}  \left( \frac{S^i}{\Nc} \right)^s \left( \frac{I^a}{\Nc} \right)^t \left( \frac{G^{ia}}{\Nc} \right)^{n - s - t},
    \end{equation}
where the coefficients $v_{stn}$ are generally momentum dependent and are at most $O(\Nc^0)$ \cite{kaplan_nucleon-nucleon_1997}.
The two-nucleon matrix element of the Hartree Hamiltonian yields the two-nucleon potential \cite{kaplan_nucleon-nucleon_1997}
    \begin{equation}
            \label{eq:largeN:potential}
        V(\vb p_+, \vb p_-) = \bra{N_\alpha (\vb p_1') N_\beta (\vb p_2') } H \ket{N_\gamma (\vb p_1) N_{\delta} (\vb p_2)} \ ,
    \end{equation}
where the Greek subscripts denote combined spin and isospin indices and 
    \begin{equation}
            \label{eq:largeN:plus_minus}
        \vb p_\pm = \vb p' \pm \vb p, 
    \end{equation}
with $\vb p' = \vb p_1' - \vb p_2'$ and $\vb p = \vb p_1 - \vb p_2$.
Also, two-nucleon matrix elements factorize in the large-\Nc limit \cite{kaplan_spin_1996},
    \begin{equation}
            \label{eq:largeN:factorize}
        \bra{N_\gamma N_\delta} \mathcal{O}_1 \mathcal{O}_2 \ket{N_\alpha N_\beta} \xrightarrow[]{N_c \to \infty} \bra{N_\gamma} \mathcal O_1 \ket{N_\alpha} \bra{N_\delta} \mathcal O_2 \ket{N_\beta} + \text{crossed},
    \end{equation}
such that the scaling of an interaction is determined from the single-nucleon matrix elements of the bilinears according to Eq.~\eqref{eq:largeN:n-body_scaling} and an overall factor of $\Nc$ is removed to account for the scaling of the Hamiltonian.

Additional $1/\Nc$ suppressions enter through the momentum dependence of the coefficients in Eq.~\eqref{eq:largeN:Hartree} \cite{kaplan_nucleon-nucleon_1997}.
In the meson exchange picture of nuclear forces, $\vb p_+$ occurs only in  relativistic corrections in t-channel diagrams and leads to inverse factors of the nucleon mass $m_N$; therefore, these contributions are suppressed since $m_N \sim O(\Nc)$.
In the u-channel, the roles of $\vb p_+$ and $\vb p_-$ are reversed and one finds identical scalings.
Therefore, it is sufficient to consider only t-channel diagrams and momenta are counted as
    \begin{equation}
            \label{eq:largeN:momenta_scaling}
        \vb p_- \sim 1 , \quad \vb p_+ \sim \Nc^{-1} \ .
    \end{equation}

In summary, in order to determine the large-\Nc scaling of various interactions, one examines the spin-isospin structure of the operators of interest and accounts for any possible suppressions due to momenta.
Then, an overall factor of \Nc is removed to account for the scaling of the Hartree Hamiltonian.
The scalings of matrix elements in the large-\Nc expansion are then mapped onto the LECs in \eftnopi with the same spin-isospin structure.

In Ref.~\cite{richardson_large-_2020}, this method was extended to include two nucleons in external magnetic and general axial fields.
This requires studying matrix elements of the form
    \begin{equation}
        \bra{N_\alpha N_\beta A} H \ket{N_\delta N_\gamma},
    \end{equation}
where $A$ is the external field and possibly carries spin and isospin indices.
By assuming that the nucleonic part of the interaction takes on a Hartree form and that two-nucleon matrix elements still factorize, the large-\Nc scalings of the LECs that couple the two-nucleon system to external fields are determined exactly as they are for \NN interactions.
Here, we treat the WIMP as an external field, and the large-\Nc constraints with an incoming and outgoing external field are obtained in the same fashion.

%% file: lagrangian.tex
\section{Large-\Nc Analysis of the Effective Lagrangian \label{sec:large_Nc_analysis}}

In this section, we derive large-\Nc constraints for the most general set of one- and two-nucleon currents in \eftnopi.
The one-nucleon operators have been derived in Refs.~\cite{fitzpatrick_effective_2013, hill_standard_2015}, and the large-\Nc constraints follow from the techniques of Refs.~\cite{dashen_1/nc_1994, dashen_spin_1995}.
The large-\Nc constraints for two-nucleon axial currents have been derived in Ref.~\cite{richardson_large-_2020}, which we reproduce here for convenience.

In principle, the currents in this work follow from a WIMP coupled to quarks and gluons inside the nucleons.
These interactions can be organized by mass dimension in Standard Model EFT (SMEFT) extended to include dark matter particles.
A complete set of operators up to and including dimension-seven have been presented in Refs.~\cite{Brod:2017bsw, Aebischer:2022wnl}.
Furthermore, these operators have been matched to the Low Energy EFT (LEFT) basis and run to hadronic scales where nonperturbative matching must be performed \cite{Crivellin:2014qxa, Hill:2014yka, hill_standard_2015, bishara_from-quarks_2017, Bishara:2018vix, Aebischer:2022wnl}.
Up to recoil corrections, the nucleon matrix elements of these operators can be parameterized through form factors that define the matching relations to couplings of the \eftnopi one-nucleon currents.
We provide the details of this matching step in Appendix~\ref{appendix:dleft}.

Large-\Nc constraints for higher dimension operators involving more than one quark bilinear can be derived, which would lead to bigger \Nc scalings than what we present in this work; however, these operators will be suppressed by powers of the SMEFT or LEFT breakdown scale.
Furthermore, if one only considers the ratios of LECs in \eftnopi, then the inclusion of higher dimension operators at the level of SMEFT or LEFT should not alter the relative scalings of the few-nucleon LECs.
This can be seen from the fact that an operator with multiple quark bilinears and the same symmetry properties of a single quark bilinear on the left hand side of Eq.~\eqref{eq:largeN:operator_expansion} only modifies the overall factor of $\Nc^m$ on the right hand side, i.e., the form of the expansion is the same up to subleading corrections.
Therefore, we implicitly present the overall large-\Nc scalings for the matrix elements of single quark bilinears but note that the \textit{relative} sizes of the LECs are independent of the interactions at the quark level.
The sizes of the WIMP-quark couplings will of course affect the overall sizes of the observables, but the combined \eftnopi and large-\Nc expansion provides model independent constraints with respect to the nuclear physics involved in the scattering process.

The matrix elements of the quark currents between two-nucleon states also admit representations in terms of form factors that can be calculated in \eftnopi. However, this also requires the inclusion of short-range operators with undetermined LECs at NLO.
This is where the greatest strength of large-\Nc constraints is found in the absence of data and lattice calculations.

\subsection{One-nucleon operators}
The one-nucleon operators have been derived in Refs.~\cite{fitzpatrick_effective_2013, hill_standard_2015}.
Here, we summarize the large-\Nc scalings of the corresponding LECs.
At leading order (LO) in \eftnopi, there are only single-nucleon zero-derivative currents, which preserve both parity ($P$) and time-reversal-invariance ($T$). 
If the WIMP is a spin-1/2 particle, then the Lagrangian at LO is 
\begin{equation}
\begin{aligned}
        \label{eq:Lagrangian:Single_Nucleon:LO}
    \calL_{\chi N}^{(PT)} & =  C_{1, \chi N}^{(PT)} \left( N^\dagger N \right) \left( \chi^\dagger \chi \right) + C_{2, \chi N}^{(PT)} \left( N^\dagger \sigma^i \tau^3 N \right) \left( \chi^\dagger \sigma^i \chi \right) \\
    & + C_{3, \chi N}^{(PT)} \left( N^\dagger \sigma^i N \right) \left( \chi^\dagger \sigma^i \chi \right) + C_{4, \chi N}^{(PT)} \left( N^\dagger \tau^3 N \right) \left( \chi^\dagger \chi \right) \, , 
\end{aligned}
\end{equation}
where $\sigma_i$ ($\tau_a$) denotes the three SU(2) Pauli matrices in spin (isospin) space.
Repeated indices are implicitly summed over.
The superscripts of the LECs indicate the properties of the corresponding operator under parity and time-reversal.
If the WIMP is a scalar particle, then there are additional operators but their contributions to elastic scattering will be suppressed by powers of the WIMP mass. 
Therefore, Eq.~\eqref{eq:Lagrangian:Single_Nucleon:LO} holds for scalar WIMPs only with the WIMP spin omitted.
Applying the large-\Nc counting rules of Eq.~\eqref{eq:largeN:none-body_scaling} shows that the LECs scale as
    \begin{align}
        C_{1, \chi N}^{(PT)}, \ C_{2, \chi N}^{(PT)} & \sim O(\Nc) \, ,\label{eq:largeN:DM-one-nucleon-LECs-Nc} \\
        C_{3, \chi N}^{(PT)}, \ C_{4, \chi N}^{(PT)} & \sim O(1) \, .
        \label{eq:largeN:DM-one-nucleon-LECs-1}
    \end{align}
Thus, the leading contributions in the combined expansion are the SI isoscalar and the SD isovector couplings.
These LECs may be recast in terms of neutron and proton couplings to the dark matter as
    \begin{align}
        C^{(PT)}_{\text{SI}, p} & = C_{1, \chi N}^{(PT)} + C_{4, \chi N}^{(PT)} \, , \\
        C^{(PT)}_{\text{SI}, n} & = C_{1, \chi N}^{(PT)} - C_{4, \chi N}^{(PT)} \, , \\
        C^{(PT)}_{\text{SD}, p} & = C_{3, \chi N}^{(PT)} + C_{2, \chi N}^{(PT)} \, , \\
        C^{(PT)}_{\text{SD}, n} & = C_{3, \chi N}^{(PT)} - C_{2, \chi N}^{(PT)} \, .
    \end{align}
Therefore, the SI proton and neutron couplings are the same up to 1/\Nc corrections while the SD couplings are of the same size but opposite sign up to 1/\Nc corrections.
The ratios of the neutron to proton couplings to $O(1/\Nc^2)$ are
    \begin{align}
        \frac{C^{(PT)}_{\text{SI}, n}}{C^{(PT)}_{\text{SI}, p}} & \approx 1 - 2 \frac{C_{4, \chi N}^{(PT)}}{C_{1, \chi N}^{(PT)}} + O(1/\Nc^2) \, , \\
        \frac{C^{(PT)}_{\text{SD}, n}}{C^{(PT)}_{\text{SD}, p}} & \approx -1 + 2 \frac{C_{3, \chi N}^{(PT)}}{C_{2, \chi N}^{(PT)}} + O(1/\Nc^2) \, .
    \end{align}
Assuming that the LECs in the $C^{(PT)}_{i, \chi N}$ basis are natural apart from their \Nc scalings, the ratios at the physical value $\Nc = 3$ are roughly 
    \begin{align}
        \abs{\frac{C^{(PT)}_{\text{SI}, n}}{C^{(PT)}_{\text{SI}, p}}} & \approx \abs{1 \pm \frac{2}{3}}  \label{eq:Laggrangian:SI_proton_neutron_ratio} \, , \\
        \abs{\frac{C^{(PT)}_{\text{SD}, n}}{C^{(PT)}_{\text{SD}, p}}} & \approx \abs{-1 \pm \frac{2}{3}} \label{eq:Laggrangian:SD_proton_neutron_ratio} \, ,
    \end{align}
where the plus and minus signs appear because the large-\Nc analysis only fixes the relative sizes of the LECs up to an overall sign.
Equations \eqref{eq:Laggrangian:SI_proton_neutron_ratio} and \eqref{eq:Laggrangian:SD_proton_neutron_ratio} imply that, in the context of isospin violating couplings, the magnitude of the ratio of the neutron to proton couplings can be between $1/3$ and $5/3$.
This range fits into the xenonphobic scenarios in which potential signals using xenon-based targets will be suppressed \cite{feng_isospin_2011, feng_xenonphobic_2013, cirigliano_shining_2014, yaguna_xenonphobic_2019}.

In the single-nucleon sector, there are one-derivative operators that violate parity and respect time-reversal invariance ($\slashed P T$) as well as operators that violate both parity and time-reversal-invariance ($\slashed P \slashed T$).
However, the forms of the operators are restricted by Galilean invariance (or invariance under infinitesimal Lorentz transformations), so the operators will contain inverse factors of $m_N$ and $m_\chi$, the nucleon mass and WIMP mass, respectively.
Therefore, the terms containing factors of $m_N$ will receive an additional $1/\Nc$ suppression while the other terms are suppressed by powers of $m_\chi$.
While we do not include these operators in the calculations of the cross sections in Sec.~\ref{sec:elastic_scattering}, we still determine the large-\Nc scalings of these currents for the sake of completeness.
The parity violating part of the Lagrangian is 
\begin{equation}
    \begin{aligned} \label{eq:Lagrangian:Single_Nucleon:P_NLO:1}
        \calL_{\chi N}^{(\slashed P T)} & =  C_{5, \chi N}^{(\slashed P T)} \epsilon^{ijk} \del^j \left( N^\dagger \sigma^k \tau^3 N \right) \left( \chi^\dagger \sigma^i \chi \right)  +  C_{6, \chi N}^{(\slashed P T)} \epsilon^{ijk} \del^j \left( N^\dagger \sigma^k N \right) \left( \chi^\dagger \sigma^i \chi \right)  \\
        & + i  C_{7, \chi N}^{(\slashed P T)} \left[ \frac{1}{2 m_N} \left( N^\dagger \Galilean^i \sigma^i \tau^3 N \right) \left( \chi^\dagger \chi \right) - \frac{1}{2 m_\chi} \left( N^\dagger \sigma^i \tau^3 N \right) \left( \chi^\dagger \Galilean^i \chi \right) \right]   \\
        & + i  C_{8, \chi N}^{(\slashed P T)} \left[ \frac{1}{2 m_N} \left( N^\dagger \Galilean^i N \right) \left( \chi^\dagger \sigma^i \chi \right) - \frac{1}{2 m_\chi} \left( N^\dagger N \right) \left( \chi^\dagger \Galilean^i \sigma^i \chi \right) \right] \\
        & + i  C_{9, \chi N}^{(\slashed P T)} \left[ \frac{1}{2 m_N} \left( N^\dagger \Galilean^i \sigma^i N \right) \left( \chi^\dagger \chi \right) - \frac{1}{2 m_\chi} \left( N^\dagger \sigma^i N \right) \left( \chi^\dagger \Galilean^i \chi \right)\right] \\
        & + i  C_{10, \chi N}^{(\slashed P T)} \left[ \frac{1}{2 m_N} \left( N^\dagger \Galilean^i \tau^3 N \right) \left( \chi^\dagger \sigma^i \chi \right) - \frac{1}{2 m_\chi} \left( N^\dagger \tau^3 N \right) \left( \chi^\dagger \Galilean^i \sigma^i \chi \right) \right] \, ,
    \end{aligned}
    \end{equation}
where $N^\dagger \Galilean^i \mathcal{O} N \equiv N^\dagger \mathcal{O}\del^i (N) - (\del^iN^\dagger) \mathcal{O}N $. The LECs scale as
    \begin{align}
        C_{5, \chi N}^{(\slashed P T)} & \sim O(\Nc) \, , \\
        C_{7, \chi N}^{(\slashed P T)}, \ C_{8, \chi N}^{(\slashed P T)} & \sim O(\Nc) \, , \\
        C_{6, \chi N}^{(\slashed P T)} & \sim O(1) \, , \\
        C_{9, \chi N}^{(\slashed P T)}, \ C_{10, \chi N}^{(\slashed P T)} & \sim O(1) \, . 
    \end{align}
Although $C_{7, \chi N}^{(\slashed P T)}$ and $C_{8, \chi N}^{(\slashed P T)}$ are $O(\Nc)$, the terms in both operators are suppressed either by the $1/\Nc$ from the nucleon mass or by a factor of $1/m_\chi$.
Therefore, both operators will yield subleading contributions to observables such as the cross section.
On the other hand, $C_{5, \chi N}^{(\slashed P T)} \sim O(\Nc)$ and is not suppressed by any factors of $m_N$ or $m_\chi$.
Moreover, the operator corresponding to $C_{5, \chi N}^{(\slashed P T)}$ is an isovector; therefore, if the WIMP-nucleon interaction is indeed isoscalar, then we would expect that the NLO corrections in the \eftnopi power counting will be highly suppressed. 
However, the fact that the only $O(\Nc)$ contribution is an isovector suggests that isospin violating dark matter should be carefully considered.

The $\slashed P \slashed T$ Lagrangian is
    \begin{equation}
        \label{eq:Lagrangian:Single_Nucleon:P_NLO:2}
    \begin{aligned}
        \calL_{\chi N}^{(\slashed P \slashed T)}  =&~  C_{11, \chi N}^{(\slashed P \slashed T)} \del^i \left( N^\dagger \sigma^i \tau^3 N \right) \left( \chi^\dagger \chi \right) + C_{12, \chi N}^{( \slashed P \slashed T)} \del^i \left( N^\dagger N \right) \left( \chi^\dagger \sigma^i \chi \right) \\
        & + C_{13, \chi N}^{( \slashed P \slashed T)} \del^i \left( N^\dagger \tau^3 N \right) \left( \chi^\dagger \sigma^i \chi \right) + C_{14, \chi N}^{(\slashed P \slashed T)} \del^i \left( N^\dagger \sigma^i N \right) \left( \chi^\dagger \chi \right)  \\
        &  + i  \epsilon^{ijk} C_{15, \chi N}^{(\slashed P \slashed T)} \left[ \frac{1}{2 m_N} \left( N^\dagger \Galilean^j \sigma^k \tau^3 N \right) \left( \chi^\dagger \sigma^i \chi \right) - \frac{1}{2 m_\chi} \left( N^\dagger \sigma^k \tau^3 N \right) \left( \chi^\dagger \Galilean^j \sigma^i \chi \right) \right] \\
        &  + i \epsilon^{ijk} C_{16, \chi N}^{(\slashed P \slashed T)} \left[ \frac{1}{2 m_N} \left( N^\dagger \Galilean^j \sigma^k N \right) \left( \chi^\dagger \sigma^i \chi \right) - \frac{1}{2 m_\chi} \left( N^\dagger \sigma^k N \right) \left( \chi^\dagger \Galilean^j \sigma^i \chi \right) \right] \, ,
        \end{aligned}
    \end{equation}
where the LECs scale as
    \begin{align}
        C_{11, \chi N}^{(\slashed P \slashed T)}, \ C_{12, \chi N}^{(\slashed P \slashed T)} & \sim O(\Nc) \, , \\
        C_{15, \chi N}^{(\slashed P \slashed T)} & \sim O(\Nc) \, \ \\ 
        C_{13, \chi N}^{(\slashed P \slashed T)}, \ C_{14, \chi N}^{(\slashed P \slashed T)} & \sim O(1) \, , \\
        C_{16, \chi N}^{(\slashed P \slashed T)} & \sim O(1) \, . 
    \end{align}
In this case, the operator proportional to $C_{15, \chi N}^{(\slashed P \slashed T)}$ is similar to those proportional to $C_{7, \chi N}^{(\slashed P \slashed T)}$ and $C_{8, \chi N}^{(\slashed P \slashed T)}$; therefore, analogous conclusions regarding the size of its contribution to observables hold.
Thus, there are only two $\slashed P \slashed T$ couplings at $O(\Nc)$; one is an isoscalar while the other is an isovector.
Isospin violating contributions can again have a significant impact on observables and should be taken into consideration in the calculations of nuclear matrix elements.

\subsection{Two-nucleon operators}
There are also two-nucleon currents that couple to dark matter.
In this work, we only consider the zero-derivative operators.
Therefore, these operators can be divided into SI or SD isoscalar, isovector, and isotensor structures.
The currents that connect two S-wave states are expected to be enhanced \cite{kaplan_new-expansion_1998, Kaplan_perturbative_1998, bedaque_effective_2002} and should contribute at NLO in \eftnopi.

It has been shown that Fierz transformations can obscure the large-\Nc scaling of the corresponding LECs, so this procedure requires caution \cite{schindler_large-$n_c$_2016}.
Specifically, if one looks only at a minimal basis of operators without considering the eliminations via Fierz transformations that produced said basis, one might conclude that a given LEC is subleading in the large-\Nc expansion when the corresponding operator may have actually arisen from the elimination of a more dominant operator.
To be concise, Fierz transformations can hide dominant large-\Nc scalings if the scalings are not properly tracked from the overcomplete set of operators.
Therefore, we use the following prescription in the passage from an overcomplete set of operators to a minimal basis.
First, an overcomplete set of operators is constructed, and the large-\Nc scaling of the matrix elements is determined through Eq.~\eqref{eq:largeN:none-body_scaling}.
Next, Fierz relations are used to eliminate the redundant operators such that the leading in large-\Nc operators are not eliminated in favor of subleading operators.
This procedure has been carried out explicitly in Refs.~\cite{richardson_large-_2020, richardson_large_2021}.

First, we consider the isoscalar contributions. 
The SI component of the Lagrangian with an overcomplete set of two-nucleon operators with no derivatives is
    \begin{equation}
        \calL^{(\text{SI}, \, s)}_{\chi NN} = \left( \chi^\dagger \chi \right) \left[ \tilde C_{1, \chi NN}^{(\text{SI}, \, s)} \left( N^\dagger N \right)^2 + \tilde C_{2, \chi NN}^{(\text{SI}, \, s)} \left( N^\dagger \sigma^i N \right)^2 + \tilde C_{3, \chi NN}^{(\text{SI}, \, s)} \left( N^\dagger \tau^a N \right)^2 + \tilde C_{4, \chi NN}^{(\text{SI}, \, s)} \left( N^\dagger \sigma^i \tau^a N \right)^2  \right] \, ,
    \end{equation}
where the superscript $s$ denote that these are isoscalar operators.
Prior to eliminating the redundant terms, the scalings of the LECs are 
    \begin{align}
        \tilde C_{1, \chi NN}^{(\text{SI}, \, s)}, \ \tilde C_{4, \chi NN}^{(\text{SI}, \, s)} & \sim O(\Nc) \ , \\
        \tilde C_{2, \chi NN}^{(\text{SI}, \, s)}, \ \tilde C_{3, \chi NN}^{(\text{SI}, \, s)} & \sim O(1/\Nc) \ .
    \end{align}
Fierz transformations relate these operators such that the minimal form of the Lagrangian is
    \begin{equation}
        \calL^{(\text{SI}, \, s)} = \left( \chi^\dagger \chi \right) \left[ C_{1, \chi NN}^{(\text{SI}, \, s)} \left( N^\dagger N \right)^2 + C_{2, \chi NN}^{(\text{SI}, \, s)} \left( N^\dagger \sigma^i N \right)^2 \right] \, ,
    \end{equation}
where the LECs are related to the overcomplete set by
    \begin{align}
        C_{1, \chi NN}^{(\text{SI}, \, s)} & = \tilde C_{1, \chi NN}^{(\text{SI}, \, s)} - 2 \tilde C_{3, \chi NN}^{(\text{SI}, \, s)} - 3 \tilde C_{4, \chi NN}^{(\text{SI}, \, s)}  \\
        C_{2, \chi NN}^{(\text{SI},s)} & = \tilde C_{2, \chi NN}^{(\text{SI}, \, s)} - \tilde C_{3, \chi NN}^{(\text{SI}, \, s)} \, ,
    \end{align}
and the LECs scale as
    \begin{align}
        C_{1, \chi NN}^{(\text{SI}, \, s)} & \sim O(\Nc) , \\
        C_{2, \chi NN}^{(\text{SI}, \, s)} & \sim O(1/\Nc) \, .
    \end{align}
This demonstrates that $\tilde C_{3, \chi N}^{(\text{SI},s)}$ constitutes a subleading correction to $C_{1, \chi N}^{(\text{SI},s)}$.
The elimination of redundancies for the remaining operators is carried out in a similar manner, which we will omit in the remainder of this section.

The SD component of the Lagrangian is 
    \begin{equation}
        \calL^{(\text{SD}, \, s)}_{\chi NN} = \left( \chi^\dagger \sigma^i \chi \right) \left[ \tilde C_{1, \chi NN}^{(\text{SD}, \, s)} \left( N^\dagger \sigma^i N \right) \left( N^\dagger N \right) + \tilde C_{2, \chi NN}^{(\text{SD}, \, s)} \left( N^\dagger \sigma^i \tau^a N \right) \left( N^\dagger \tau^a N \right) \right] \, .
    \end{equation}
However, these operators are not linearly independent and they both lead to the same scaling in large-\Nc.
Therefore, the Lagrangian can be written as
    \begin{equation}
        \calL^{(\text{SD}, \, s)}_{\chi NN} = C_{1, \chi NN}^{(\text{SD}, \, s)} \left( \chi^\dagger \sigma^i \chi \right) \left( N^\dagger \sigma^i N \right) \left( N^\dagger N \right) \, ,
    \end{equation}
where 
    \begin{equation}
        C_{1, \chi NN}^{(\text{SD}, \, s)} \sim O(1) \, .
    \end{equation}

Now, we turn to the isovector operators. 
The SI part of the Lagrangian is
    \begin{equation}
        \calL^{(\text{SI}, \, v)}_{\chi NN} = \left( \chi^\dagger  \chi \right) \left[\tilde C_{1, \chi NN}^{(\text{SI}, \, v)} \left( N^\dagger \tau^3 N \right) \left( N^\dagger N \right) + \tilde C_{2, \chi NN}^{(\text{SI}, \, v)} \left( N^\dagger \sigma^i \tau^3 N \right) \left( N^\dagger \sigma^i N \right) \right] \, ,
    \end{equation}
where the superscript $v$ indicates isovector contributions and
    \begin{equation}
       \tilde C_{1, \chi NN}^{(\text{SI}, \, v)},~ \tilde C_{2, \chi NN}^{(\text{SI}, \, v)} \sim O(1) \, .
    \end{equation}
Again, the operators are not independent and they both lead to the same large-\Nc scaling.
We choose to retain the first operator such that the Lagrangian is
    \begin{equation}
        \calL^{(\text{SI}, \, v)}_{\chi NN} = C_{1, \chi NN}^{(\text{SI}, \, v)} \left( \chi^\dagger \chi \right)  \left( N^\dagger \tau^3 N \right) \left( N^\dagger N \right) \, .
    \end{equation}
The SD isovector terms are 
    \begin{equation}
        \begin{aligned}
            \calL^{(\text{SD}, \, v)}_{\chi NN} & =  \left( \chi^\dagger \sigma^i \chi \right) \bigg[ \tilde C_{1, \chi NN}^{(\text{SD}, \, v)} \left( N^\dagger \sigma^i \tau^3 N \right) \left( N^\dagger N \right) + \tilde C_{2, \chi NN}^{(\text{SD}, \, v)} \left( N^\dagger \sigma^i N \right) \left( N^\dagger \tau^3 N \right)  \\
            &  ~+ \tilde C_{3, \chi NN}^{(\text{SD}, \, v)} \epsilon^{ijk} \epsilon^{3ab} \left( N^\dagger \sigma^j \tau^a N \right) \left( N^\dagger \sigma^k \tau^b N \right) + \tilde C_{4, \chi NN}^{(\text{SD}, \, v)} \epsilon^{ijk} \left( N^\dagger \sigma^j \tau^3 N \right) \left( N^\dagger \sigma^k N \right) \\
            & ~+ \tilde C_{5, \chi NN}^{(\text{SD}, \, v)} \epsilon^{3ab} \left( N^\dagger \sigma^i \tau^a N \right) \left( N^\dagger \tau^b N \right) \bigg]\,.
        \end{aligned}
    \end{equation}
The first three terms conserve both parity and time-reversal-invariance, but the last two terms conserve parity and violate time-reversal-invariance.
There are only two independent operators; in particular, there is one $T$ conserving operator and one $\slashed T$ operator.
We choose to retain
    \begin{equation}
        \calL^{(\text{SD}, \, v)}_{\chi NN} = \left( \chi^\dagger \sigma^i \chi \right) \left[ C_{1, \chi NN}^{(\text{SD}, \, v)} \epsilon^{ijk} \epsilon^{3ab} \left( N^\dagger \sigma^j \tau^a N \right) \left( N^\dagger \sigma^k \tau^b N \right) +  C_{2, \chi NN}^{(\text{SD}, \, v)} \epsilon^{ijk} \left( N^\dagger \sigma^j \tau^3 N \right) \left( N^\dagger \sigma^k N \right) \right] \, ,
    \end{equation}
where 
    \begin{align}
        C_{1, \chi NN}^{(\text{SD}, \, v)} & \sim O(\Nc) \, , \\
        C_{2, \chi NN}^{(\text{SD}, \, v)} & \sim O(1) \, .
    \end{align}
Thus, the SD isoscalar and SI isovector $PT$ couplings are 1/\Nc suppressed relative to the SD isovector coupling. 
The $P\slashed T$ term is also 1/\Nc suppressed relative to the symmetry conserving term.
    
Lastly, we examine the isotensor operators.
There are two possible SI operators,
\begin{equation}
    \begin{aligned}
        \calL^{(\text{SI}, \, t)}_{\chi NN} & = \left( \chi^\dagger \chi \right) \left\{ \tilde C_{1, \chi NN}^{(\text{SI}, \, t)} \left[ \left( N^\dagger \sigma^i \tau^3 N \right) \left( N^\dagger \sigma^i \tau^3 N \right) - \frac{1}{3} \left( N^\dagger \sigma^i \tau^a N \right) \left( N^\dagger \sigma^i \tau^a N \right) \right] \right. \\
        & \left. + \tilde C_{2, \chi NN}^{(\text{SI}, \, t)} \left[ \left( N^\dagger \tau^3 N \right) \left( N^\dagger \tau^3 N \right) - \frac{1}{3} \left( N^\dagger \tau^a N \right) \left( N^\dagger \tau^a N \right) \right] \right\} \,,
    \end{aligned}
    \end{equation}
where $t$ denotes isotensor contributions.
The second term in each set of square brackets is present in order to isolate the $I=2$ representation. 
Fierz transformations give only one independent operator,
    \begin{equation}
        \calL^{(\text{SI}, \, t)}_{\chi NN} =  C_{1, \chi NN}^{(\text{SI}, \, t)} \left( \chi^\dagger \chi \right) \left[ \left( N^\dagger \sigma^i \tau^3 N \right) \left( N^\dagger \sigma^i \tau^3 N \right) -\frac{1}{3} \left( N^\dagger \sigma^i \tau^a N \right) \left( N^\dagger \sigma^i \tau^a N \right) \right] \,,
    \end{equation}
where 
    \begin{equation}
        C_{1, \chi NN}^{(\text{SI}, \, t)} \sim O(\Nc) \, .
    \end{equation}
We could in principle include a spin dependent operator of the form
    \begin{equation}
        \left( \chi^\dagger \sigma^i \chi \right) \left[ \left( N^\dagger \sigma^i \tau^3 N \right) \left( N^\dagger \tau^3 N \right) - \left( N^\dagger \sigma^i \tau^a N \right) \left( N^\dagger \tau^a N \right) \right] \, .
    \end{equation}
However, Fierz transformations show that this operator vanishes.

In summary, we have seven independent operators in total and the complete Lagrangian is 
    \begin{align}
            \label{eq:Lagrangian:two_nucleon_minimal}
        \calL_{\chi,NN} & =  C_{1, \chi NN}^{(\text{SI}, \, s)} \left( \chi^\dagger \chi \right)  \left( N^\dagger N \right) \left( N^\dagger N \right)  + C_{2, \chi NN}^{(\text{SI}, \, s)} \left( \chi^\dagger \chi \right)  \left( N^\dagger \sigma^i N \right) \left( N^\dagger \sigma^i N \right)  \nonumber \\
        &   + C_{1, \chi NN}^{(\text{SD}, \, s)} \left( \chi^\dagger \sigma^i \chi \right)  \left( N^\dagger \sigma^i N \right) \left( N^\dagger N \right)  + C_{1, \chi NN}^{(\text{SI}, \, v)} \left( \chi^\dagger \chi \right)  \left( N^\dagger \tau^3 N \right) \left( N^\dagger N \right)  \nonumber \\
        &   + \epsilon^{ijk} \epsilon^{3ab} C_{1, \chi NN}^{(\text{SD}, \, v)} \left( \chi^\dagger \sigma^i \chi \right) \left( N^\dagger \sigma^j \tau^a N \right) \left( N^\dagger \sigma^k \tau^b N \right) + \epsilon^{ijk} C_{2, \chi NN}^{(\text{SD}, \, v)}  \left( \chi^\dagger \sigma^i \chi \right) \left( N^\dagger \sigma^j \tau^3 N \right) \left( N^\dagger \sigma^k N \right)  \nonumber \\
        & +  C_{1, \chi NN}^{(\text{SI}, \, t)} \left( \chi^\dagger \chi \right) \left[ \left( N^\dagger \sigma^i \tau^3 N \right) \left( N^\dagger \sigma^i \tau^3 N \right) - \frac{1}{3} \left( N^\dagger \sigma^i \tau^a N \right) \left( N^\dagger \sigma^i \tau^a N \right) \right] ,
    \end{align}
where
    \begin{align}
        C_{1, \chi NN}^{(\text{SI}, \, s)}, C_{1, \chi NN}^{(\text{SD}, \, v)}, C_{1, \chi NN}^{(\text{SI}, \, t)} & \sim O(\Nc) \, , \\
        C_{1, \chi NN}^{(\text{SD}, \, s)}, C_{1, \chi NN}^{(\text{SI}, \, v)}, C_{2, \chi NN}^{(\text{SD}, \, v)} & \sim O(1) \, , \\
        C_{2, \chi NN}^{(\text{SI}, \, s)} & \sim O(1/\Nc) \, .
    \end{align}
Therefore, the isoscalar and isotensor SI LECs are of the same size, so we emphasize again that isospin violating WIMP-nucleus interactions should be carefully considered even in the SI case.
Additionally, the $P \slashed T$ coupling $C_{2, \chi NN}^{(\text{SD}, \, v)}$ is 1/\Nc suppressed.

It is often convenient to work with operators where the nucleonic portion is written in terms of partial waves.
Because the operators in Eq.~\eqref{eq:Lagrangian:two_nucleon_minimal} do not contain derivatives, we only retain the operators that connect two-nucleon S-wave states.
Using Fierz transformations to translate Eq.~\eqref{eq:Lagrangian:two_nucleon_minimal} into the partial wave basis results in
\begin{equation}
    \begin{aligned}
     \label{eq:Lagrangian:two_nucleon_partial}
        \calL_{\chi NN}  = &~  2 \left( C_{1, \chi NN}^{(\text{SI}, \, s)} + C_{2, \chi NN}^{(\text{SI}, \, s)} \right) \left( \chi^\dagger \chi \right) \left[ \left( N^T P^i N \right)^\dagger \left( N^T P^i N \right) \right] \\
        & + 2 \left( C_{1, \chi NN}^{(\text{SI}, \, s)} - 3 C_{2, \chi NN}^{(\text{SI}, \, s)} \right) \left( \chi^\dagger \chi \right) \left[ \left( N^T \bar P^a N \right)^\dagger \left( N^T \bar P^a N \right) \right]  \\
        &  + 12 C_{1, \chi NN}^{(\text{SI}, \, t)} \left( \chi^\dagger \chi \right) \left[ \left( N^T \bar P^3 N \right)^\dagger \left( N^T \bar P^3 N \right) - \frac{1}{3} \left( N^T \bar P^a N \right)^\dagger \left( N^T \bar P^a N \right)  \right]  \\
        &  + 8 C_{1, \chi NN}^{(\text{SD}, \, v)} \left( \chi^\dagger \sigma^i \chi \right) \left[ \left( N^T \bar P^3 N \right)^\dagger \left( N^T P^i N \right) + \text{H.c.} \right] \\
        &- 2i \epsilon^{ijk} C_{1, \chi NN}^{(\text{SD}, \, s)} \left( \chi^\dagger \sigma^i \chi \right) \left( N^T P^j N \right)^\dagger \left( N^T P^k N \right)  \\
        &  - 2i \epsilon^{3ab} C_{1, \chi NN}^{(\text{SI}, \, v)} \left( \chi^\dagger \chi \right) \left( N^T \bar P^a N \right)^\dagger \left( N^T \bar P^b N \right)  \\
        & + 4  C_{2, \chi NN}^{(\text{SD}, v)}\left( \chi^\dagger \sigma^i \chi \right) \left[ i \left( N^T P^i N \right)^\dagger \left( N^T \bar P^3 N \right) + \text{H.c.} \right] \, ,
    \end{aligned} 
    \end{equation}
where $P^i =\sigma_2\sigma_i\tau_2/\sqrt{8}$ and $\bar{P}^a =\sigma_2\tau_2\tau_a/\sqrt{8}$ are the projection operators onto the spin triplet-isospin singlet and spin singlet-isospin triplet states, respectively.

%% file: calculations/calculations_main.tex
\section{Dark Matter-Nucleus Elastic Scattering}
    \label{sec:elastic_scattering}

Before presenting the results for specific targets, we briefly review the calculation of the WIMP-nucleus cross section.
The differential scattering cross section for dark matter scattering off of a nucleus initially at rest is related to the scattering amplitude $\calM$ through \cite{del_nobile_tools_2013} 
    \begin{equation}
            \label{eq:Elastic:diff_cross}
        \frac{d \sigma}{d E_R} = \frac{1}{32 \pi m_\chi^2 m_T v_\chi^2} \abs{\calM}^2 \, ,
    \end{equation}
where $E_R$ is the recoil energy, $m_T$ is the mass of the target nucleus, and $v_\chi$ is the dark matter velocity in the lab frame.
Since we take the nucleus to be at rest initially, the recoil energy can be expressed as $E_R = q^2/2 m_T$, where $q$ is the momentum transfer.
In order to draw a comparison with Refs.~\cite{korber_first-principle_2017,Andreoli:2018etf}, we define response functions $F_i^{(\nu)}(q^2)$ through
    \begin{align}
            \label{eq:response}
        \abs{\calM_A (q^2)}^2 & = (2 m_T)^2 (2 m_\chi)^2 \sigma^{\text{SI}}_{0,N} \frac{\pi A^2}{m_{\chi N}^2} \abs{\sum_{i, \nu} \alpha_i F_{i}^{(\nu)}(q^2)}^2 \, ,
    \end{align}
where $\nu$ is the order in the EFT expansion, $\alpha$ is a generic coupling of type $i$ (e.g. SI or SD, isoscalar or isovector) that can be related to couplings at the quark level, $m_{\chi N}$ is the WIMP-nucleon reduced mass, and $\sigma^{\text{SI}}_{0,N}$ is the single nucleon isoscalar cross section at zero momentum transfer.
When we consider the response functions below, we will express $\sigma^{\text{SI}}_{0,N}$ in terms of the pion-nucleon sigma term (see Appendix \ref{appendix:dleft}).

In the following sections, we calculate the unpolarized cross sections; thus, the amplitude in Eqs.~\eqref{eq:Elastic:diff_cross} and \eqref{eq:response} should also include the average over incoming spins and the sum over outgoing spins.
Then, we will provide the relative scalings of various cross sections at zero momentum transfer with respect to \Nc. 
For this, it is sufficient to work to LO in the combined large-\Nc and \eftnopi expansion as subleading corrections should not drastically alter the relative sizes.
However, when we study the response functions of the deuteron, ${}^3$H, and ${}^3$He that arise from the quark scalar current we will consider the NLO corrections in the \eftnopi power counting and use the two-body large-\Nc constraints to set the sizes of the LECs.

\input{calculations/nucleon_amplitude}

\input{calculations/deuteron_amplitude}

\input{calculations/triton_amplitude}

%% file: calculations/nucleon_amplitude.tex
\subsection{Dark Matter-Nucleon Elastic Scattering}
    \label{sec:Nucleon_elastic}

At LO in the \eftnopi expansion, the WIMP-nucleon elastic scattering amplitude only receives contributions from Eq.~\eqref{eq:Lagrangian:Single_Nucleon:LO}.
Consider a nucleon initially at rest with spin $\nu$ and isospin $b$, an outgoing nucleon with spin $\mu$ and isospin $a$, and an incoming (outgoing) dark matter particle with spin $r$ ($s$) and momentum $\vb k$ ($\vb k'$).
The momentum of the outgoing nucleon will be $\vb q = \vb k - \vb k'$.
The unpolarized scattering amplitude can be expressed as
    \begin{equation}
        \frac{1}{4} \sum_{\text{spins}} \abs{\calM}^2 = 16 m_N^2 m_\chi^2 \left[ \left( C_{1, \chi N}^{(PT)} \pm C_{4, \chi N}^{(PT)} \right)^2 + 3 \left( C_{2, \chi N}^{(PT)} \pm C_{3, \chi N}^{(PT)} \right)^2 \right] \, ,
    \end{equation}
where the upper signs correspond to proton-dark matter scattering and lower signs correspond to neutron-dark matter scattering.
Additionally, the first term arises from the SI interactions while the second term arises from the SD interactions.
We find
    \begin{align}
        \left( C^{(PT)}_{1, \chi N} \pm C^{(PT)}_{4, \chi N} \right)^2 = \frac{\pi}{m_{\chi N}^2} \sigma^{\text{SI}}_{0, \, p/n} F^2_{\text{SI}, \, p/n} (q^2) \, , \\
        \left( C^{(PT)}_{2, \chi N} \pm C^{(PT)}_{3, \chi N} \right)^2 = \frac{\pi}{3 m_{\chi N}^2} \sigma^{\text{SD}}_{0, \, p/n} F^2_{\text{SD}, \, p/n} (q^2) \, .
    \end{align}
By definition, the form factors approach 1 as $q^2 \to 0$.
Additionally, the operators in Eqs.~\eqref{eq:Lagrangian:Single_Nucleon:P_NLO:1} and \eqref{eq:Lagrangian:Single_Nucleon:P_NLO:2} introduce explicit momentum dependence and their contributions will vanish in this limit.
Therefore, at $q^2 = 0$ we have
    \begin{align}
        \left( C^{(PT)}_{1, \chi N} \pm C^{(PT)}_{4, \chi N} \right)^2 = \frac{\pi}{m_{\chi N}^2} \sigma^{\text{SI}}_{0, \, p/n} \, , \\
        \left( C^{(PT)}_{2, \chi N} \pm C^{(PT)}_{3, \chi N} \right)^2 = \frac{\pi}{3 m_{\chi N}^2} \sigma^{\text{SD}}_{0, \, p/n} \, ,
    \end{align}
which is exact.
Thus, at \LONc the ratio of the cross sections is
    \begin{equation}
        \frac{\sigma^{\text{SI}}_{0,N}}{\sigma^{\text{SD}}_{0,N}} = \frac{C^{(PT) 2}_{1, \chi N} \left( 1 \pm \frac{C^{(PT)}_{4, \chi N}}{C^{(PT)}_{1, \chi N}} \right)^2 }{3 C^{(PT) 2}_{2, \chi N} \left( 1 \pm \frac{C^{(PT)}_{3, \chi N}}{C^{(PT)}_{2, \chi N}} \right)^2} \approx \frac{C^{(PT) 2}_{1, \chi N}}{3 C^{(PT) 2}_{2, \chi N}} \sim \frac{1}{3} \, ,
    \end{equation}
for all proton and neutron combinations if both $C_{1, \chi N}^{(PT)}$ and $C_{2, \chi N}^{(PT)}$ are considered to be of natural size apart from their $\Nc$ scalings.
We expect this ratio to receive roughly 30$\%$ corrections at NLO in the large-\Nc expansion as well as \eftnopi corrections.

Alternatively, the ratio of the SI proton cross section to the SI neutron cross section is
    \begin{equation}
            \label{eq:SI_cross_p/n}
        \frac{\sigma^{\text{SI}}_{0,p}}{\sigma^{\text{SI}}_{0,n}} = \frac{\left( 1 + \frac{C^{(PT)}_{4, \chi N}}{C^{(PT)}_{1, \chi N}} \right)^2}{\left( 1 - \frac{C^{(PT)}_{4, \chi N}}{C^{(PT)}_{1, \chi N}} \right)^2} \approx 1 + 4 \frac{C^{(PT)}_{4, \chi N}}{C^{(PT)}_{1, \chi N}} + O(1/\Nc^2) \, .
    \end{equation}
Therefore, at LO in the large-\Nc expansion the SI cross sections for the proton and neutron are roughly the same size.
Recall $C^{(PT)}_{4, \chi N}/C^{(PT)}_{1, \chi N} \sim O(1/\Nc)$, but this ratio on the right-hand side is multiplied by an additional factor of 4.
Thus, at the physical value $\Nc=3$, the second term can lead to significant cancellations or enhancements depending on the relative signs of the LECs.
Similarly, the ratio of the SD cross sections is
    \begin{equation}
            \label{eq:SD_cross_p/n}
        \frac{\sigma^{\text{SD}}_{0,p}}{\sigma^{\text{SD}}_{0,n}} = \frac{\left( 1 + \frac{C^{(PT)}_{3, \chi N}}{C^{(PT)}_{2, \chi N}} \right)^2}{\left( 1 - \frac{C^{(PT)}_{3, \chi N}}{C^{(PT)}_{2, \chi N}} \right)^2} \approx 1 + 4 \frac{C^{(PT)}_{3, \chi N}}{C^{(PT)}_{2, \chi N}} + O(1/\Nc^2) \, ,
    \end{equation}
from which the same conclusions regarding the SI cross sections may be drawn.

%% file: calculations/deuteron_amplitude.tex
\subsection{WIMP-Deuteron Elastic Scattering}
    \label{sec:Deuteron}
    
\vspace{-0.58in}

\begin{center}
    \begin{figure}
        \centering
        \includegraphics[scale=1.2]{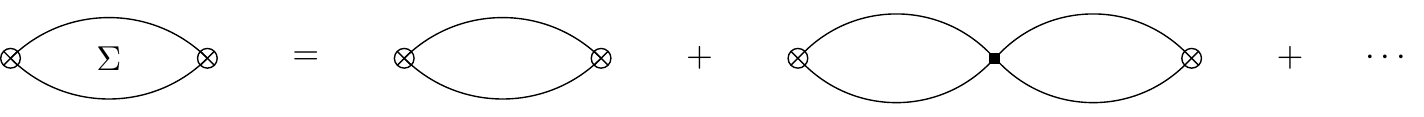}
        \caption{Irreducible two-point function. The crossed circles represent the interpolating deuteron field, the solid lines represent nucleons, and the black square represents the two-derivative two-nucleon vertex. The ellipsis denotes higher order terms.}
        \label{fig:deuteron_2pt}
    \end{figure}
\end{center}

\begin{center}
    \begin{figure}
        \centering
        \includegraphics[scale=1.3]{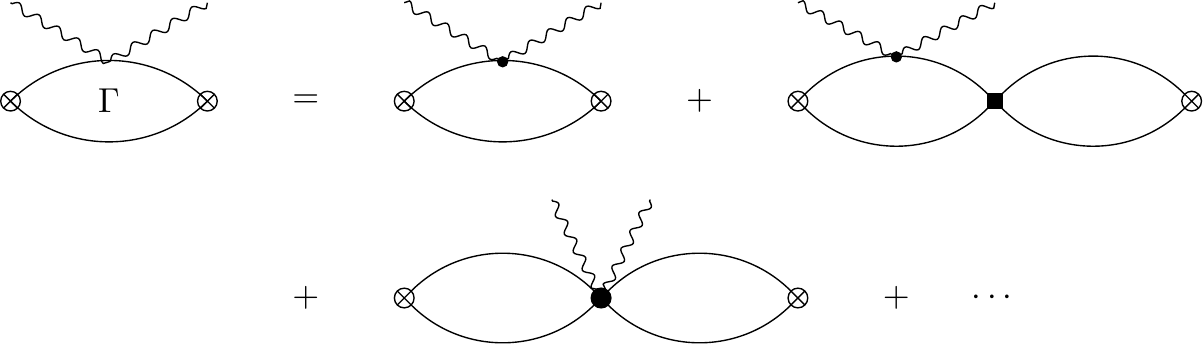}
        \caption{Irreducible four-point function for elastic WIMP-deuteron scattering. The wavy lines represent the WIMP. The small black dot represents the zero-derivative WIMP-one-nucleon coupling, and the large black circle represents the zero-derivative WIMP-two-nucleon coupling. The ellipsis denotes higher order terms.}
        \label{fig:wimp_deuteron_4pt}
    \end{figure}
\end{center}
    
Now, we consider parity and time-reversal invariant WIMP-deuteron elastic scattering without considering a specific model.
First, we review the technology developed in Ref.~\cite{Kaplan_perturbative_1998} for calculating amplitudes involving the deuteron. 
Additionally, we will need the $\NN$ \eftnopi Lagrangian at NLO, \cite{chen_nucleon-nucleon_1999}
    \begin{align}
            \label{eq:NN_Lagrangian}
        \calL_{NN} =&~ \Ndag \lb i\partial_0 +\frac{\grad^2}{2 m_N}\rb N - C^{(\threeS)}_0(N^TP^iN)^\dagger(N^TP^iN)- C^{(\oneS)}_0(N^T\bar P^aN)^\dagger(N^T\bar P^aN)  \nonumber \\
        &+ \frac{C^{(\threeS)}_2}{8}\left[(N^TP^i N)^\dagger(N^T P^i\Galilean^2 N) + \text{H.c}\right]+ \frac{C^{(\oneS)}_2}{8}\left[(N^T \bar P^aN)^\dagger (N^T \bar P^a\Galilean^2 N) + \text{H.c}\right]  \, .
    \end{align}
Using PDS \cite{kaplan_new-expansion_1998} and the $Z$ parameterization \cite{Phillips:1999hh}, the couplings $C_{2n}^{(s)}$, where $2n$ is the number of derivatives in the operator and $s$ is the partial wave channel, can be used to reproduce the location of the deuteron pole and its residue.
Elements of the S-matrix and thus the scattering amplitude are related to correlation functions through the LSZ formula
    \begin{equation}
            \label{eq:Deuteron:LSZ}
        \bra{p', m; k', r} S \ket{p, n; k, s} = i \left[ \frac{ \Gamma (q^2, \bar E, \bar E') }{ d \Sigma(\bar E)/dE } \right]_{\bar E, \bar E' \to -B} \, ,
    \end{equation}
where $\Sigma$ is the irreducible two-point function in Fig.~\ref{fig:deuteron_2pt}, $\Gamma$ is the irreducible four-point correlation function in Fig.~\ref{fig:wimp_deuteron_4pt}, $p$ ($p'$) is the momentum of the incoming (outgoing) deuteron, $m$ ($n$) is the polarization of the incoming (outgoing) deuteron, $k$ ($k'$) is the momentum of the incoming (outgoing) WIMP, $q$ is the momentum transfer, and $s$ ($r$) is the incoming (outgoing) spin index for a spin-1/2 WIMP.
The two-nucleon center-of-mass energy $\bar E$ is given by
    \begin{equation}
        \bar E = E - \frac{p^2}{4 m_N} \, ,
    \end{equation}
where the energy of each nucleon is $E/2$.
An analogous relationship holds for the energy of the outgoing deuteron $E'$.

The functions $\Sigma$ and $\Gamma$ are expanded in powers of $Q$,
    \begin{align}
        \Sigma \left( \bar E \right) & = \sum_{n=1} \Sigma_{(n)} \left( \bar E \right) \, , \\
        \Gamma \left( q^2, \bar E, \bar E' \right) & = \sum_{n=-1} \Gamma_{(n)} \left( q^2, \bar E, \bar E' \right) \, ,
    \end{align}
where $n$ denotes each term's order in the $Q/\LambdaNoPion$ expansion.
In PDS, the functions for $\Sigma_1$ and $\Sigma_2$ are \cite{Kaplan_perturbative_1998}
    \begin{align}
        \Sigma_{(1)} (\bar E) & = - \frac{i m_N}{4 \pi} \left( \mu - \sqrt{-m_N \bar E - i \epsilon} \right) \, , \\
        \Sigma_{(2)} (\bar E) & = -i C_2^{(\threeS)} m_N \bar E \Sigma_{(1)}^2 (\bar E) \, ,
    \end{align}
where $\mu$ is the renormalization scale.
To NLO, Eq.~\eqref{eq:Deuteron:LSZ} is
    \begin{equation}
        \bra{p', m; k', r} S \ket{p, n; k, s} = i \left[ \frac{\Gamma_{(-1)}}{d \Sigma_{(1)} / d \bar E} + \frac{\Gamma_{(0)} \left( d \Sigma_{(1)} / d \bar E \right) - \Gamma_{(-1)} \left( d \Sigma_{(2)} / d \bar E \right)}{ \left( d \Sigma_{(1)} / d \bar E \right)^2 } + \cdots \right] \, .
    \end{equation}

Now, we must calculate the contributions from the irreducible four-point diagrams. 
The LO contributions consist of insertions of the isoscalar operators in Eq.~\eqref{eq:Lagrangian:Single_Nucleon:LO} yielding
    \begin{eqnarray}
        \Gamma_{(-1)} (q^2, E, E') & = & \frac{m_N^2}{\pi q} \left[ i \delta^{rs} \delta^{mn} C^{(PT)}_{1, \chi N} + \sigma^i_{rs} \epsilon^{imn} C^{(PT)}_{3, \chi N} \right] \tan^{-1} \left( \frac{q}{ 4 \gamma_t } \right) \, ,
    \end{eqnarray}
where $\gamma_t$ is the deuteron binding momentum.

At NLO, there are contributions from the two-derivative two-nucleon operator proportional to $C_2^{(\threeS)}$ in Eq.~\eqref{eq:NN_Lagrangian} with an insertion of the single-nucleon currents in Eq.~\eqref{eq:Lagrangian:Single_Nucleon:LO} giving
    \begin{eqnarray}
        \Gamma_{(0),1b} & = &  - \frac{2 m_N^3}{ (4 \pi)^2 } C_2^{(\threeS)} \left( i C^{(PT)}_{1, \chi N} \delta^{mn} \delta^{rs} + C^{(PT)}_{3, \chi N} \epsilon^{imn} \sigma^i_{rs} \right) \left( \mu - \gamma_t \right) \left[ \mu - \gamma_t - \frac{4 \gamma_t^2}{q} \tan^{-1} \left( \frac{q}{4 \gamma_t} \right) \right] \, .
    \end{eqnarray}
Finally, there are two contributions at NLO from the WIMP-two-nucleon contact terms in Eq.~\eqref{eq:Lagrangian:two_nucleon_partial} that connect two $\threeS$ $NN$ states leading to
    \begin{equation}
        \Gamma_{(0),2b} = - \left( \frac{m_N}{4 \pi} \right)^2 \left( \mu - \gamma_t \right)^2 \left[ -2i \left( C_{1, \chi NN}^{(\text{SI}, \, s)} + C_{2, \chi NN}^{(\text{SI}, \, s)} \right) \delta^{mn} \delta^{rs} - 2 \epsilon^{imn} C_{1, \chi NN}^{(\text{SD}, \, s)} \sigma^i_{rs} \right] \, . 
    \end{equation}
Thus, the amplitude up to and including NLO is given by
    \begin{align}
        i \calM & = 4 m_d m_\chi \left\{ - \frac{8 \gamma_t}{q} \tan^{-1} \left( \frac{q}{ 4 \gamma_t } \right)  \left[ i \delta^{rs}  \delta^{mn} C^{(PT)}_{1, \chi N} \nonumber + \sigma^i_{rs} \epsilon^{imn} C^{(PT)}_{3, \chi N} \right] \right. \nonumber \\
        & + \frac{m_N \gamma_t}{\pi} C_2^{(\threeS)} \left[ i \delta^{rs}  \delta^{mn} C^{(PT)}_{1, \chi N} \nonumber + \sigma^i_{rs} \epsilon^{imn} C^{(PT)}_{3, \chi N} \right] \left( \mu - \gamma_t \right)^2 \left[ 1 - \frac{4 \gamma_t}{q} \tan^{-1} \left( \frac{q}{4 \gamma_t} \right) \right] \nonumber \\
        & \left. - \frac{\gamma_t}{\pi} \left( \mu - \gamma_t \right)^2 \left[ i \left( C_{1, \chi NN}^{(\text{SI}, \, s)} + C_{2, \chi NN}^{(\text{SI}, \, s)} \right) \delta^{mn} \delta^{rs} +  \epsilon^{imn} C_{1, \chi NN}^{(\text{SD}, \, s)} \sigma^i_{rs} \right] \right\} \, .
    \end{align}
The $\mu$ dependence of the WIMP-two-nucleon couplings is determined by requiring the amplitude to be independent of $\mu$.
Because the first two terms in the amplitude are already $\mu$ independent, the third term should separately $\mu$ independent.
Therefore, the LECs depend on the PDS subtraction point according to
    \begin{equation}
        C_{1, \chi NN}^{(\text{SI}, \, s)} + C_{2, \chi NN}^{(\text{SI}, \, s)}, \ C_{1, \chi NN}^{(\text{SD}, \, s)} \propto \left( \mu - \gamma_t \right)^{-2} \, ,
    \end{equation}
so the familiar infrared enhancement occurs when $\mu \sim Q$.
It has been discussed that large-\Nc constraints for two-nucleon contact terms conflict with the renormalization group for certain values of $\mu < m_\pi$ \cite{kaplan_spin_1996,schindler_large-$n_c$_2018,mehen_wigner_1999,Calle_2008,Calle_2009,Ruiz_2016}.
However, Ref.~\cite{Epelbaum_wilsonian_RG_2017} suggests that it is possible to take $\mu > m_\pi$ and still obtain a consistent power counting.

While renormalization group invariance fixes the $\mu$ dependence of the LECs, other dimensional factors are required to produce the correct mass dimension of the LECs.
In the case that the interactions are mediated by some heavy exchange particle, it is expected that two of these factors will come from the mass of the mediator.
In the absence of other scales, factors of $m_\pi$ are included to obtain the correct mass dimension, in specific cases such as the two-nucleon axial coupling the additional factors might also include the nucleon mass.
Moreover, factors of $4 \pi$ generically arise from loop diagrams that also enhance the LECs.
All together, we expect the power counting of the two-body LECs to be
    \begin{equation}
        C_{i, \, \chi \NN} \sim \frac{4 \pi}{\Lambda^2 Q^2 M} \, ,
    \end{equation}
where $\Lambda$ is on the order of the WIMP-quark mediator mass, and $M$ is on the order of either the pion or nucleon mass.

Next, we obtain the SI and SD cross sections.
Averaging over initial spins and summing over final spins in the squared amplitude yields
    \begin{align}
        \frac{1}{6} \sum_{\text{spins}} \abs{\calM}^2 & = 16 m_d^2 m_\chi^2 \left\{ \left[ 2 C_{1, \chi N}^{(PT)} \left( \frac{4 \gamma_t}{q} \tan^{-1} \left( \frac{q}{4 \gamma_t} \right) + \frac{m_N \gamma_t}{2 \pi} C_2^{(\threeS)} \left( \mu - \gamma_t \right)^2 \right. \right. \right. \nonumber \\
        & \left. \left. \left.- \frac{2 m_N \gamma_t^2}{\pi q} C_2^{(\threeS)} \left( \mu - \gamma_t \right)^2 \tan^{-1} \left( \frac{q}{4 \gamma_t} \right) \right)  \right. \right. \left. \left.  -  \frac{\gamma_t}{\pi} \left( \mu - \gamma_t \right)^2 \left( C_{1, \chi NN}^{\left( \text{SI}, \, s \right)} + C_{2, \chi NN}^{\left( \text{SI}, \, s \right)} \right)  \right]^2   \right. \nonumber \\
        &  +  2  \left. \left[ 2 C_{3, \chi N}^{(PT)} \left( \frac{4 \gamma_t}{q} \tan^{-1} \left( \frac{q}{4 \gamma_t} \right) + \frac{m_N \gamma_t}{2 \pi} C_2^{(\threeS)} \left( \mu - \gamma_t \right)^2  - \frac{2 m_N \gamma_t^2}{\pi q} C_2^{(\threeS)} \left( \mu - \gamma_t \right)^2 \tan^{-1} \left( \frac{q}{4 \gamma_t} \right) \right) \right. \right. \nonumber \\
        & \left. \left. - \frac{\gamma_t}{\pi} \left( \mu - \gamma_t \right)^2  C_{1, \chi NN}^{\left( \text{SD}, \, s \right) } \right]^2   \right\}\,. 
    \end{align}
Therefore, the cross sections in the limit $q^2 \to 0$ are
    \begin{align}
        \sigma^{\text{SI}}_{0,d} & = \frac{m_{\chi d}^2}{\pi} \left[ 2 C_{1, \chi N}^{(PT)} - \frac{\gamma_t}{\pi} \left( \mu - \gamma_t \right)^2 \left( C_{1, \chi NN}^{\left( \text{SI}, \, s \right)} + C_{2, \chi NN}^{\left( \text{SI}, \, s \right)} \right) \right]^2 \, , \label{eq:cross_section:deuteron_SI}\\
        \sigma^{\text{SD}}_{0,d} & = \frac{2 m_{\chi d}^2}{\pi} \left[ 2 C_{3, \chi N}^{(PT)}  - \frac{\gamma_t}{\pi} \left( \mu - \gamma_t \right)^2  C_{1, \chi NN}^{\left( \text{SD}, \, s \right) } \right]^2 \,  . \label{eq:cross_section:deuteron_SD}
    \end{align}
The form factors are given by
    \begin{align}
        F_{\text{SI}}(q^2) & = \frac{4 \gamma_t}{q} \tan^{-1} \left( \frac{q}{4 \gamma_t} \right) - \frac{m_N \gamma_t}{2 \pi} C_2^{(\threeS)} \left( \mu - \gamma_t \right)^2 \left[ 1 - \frac{4 \gamma_t}{q} \tan^{-1} \left( \frac{q}{4 \gamma_t} \right) \right] + \frac{\gamma_t}{\pi} \left( \mu - \gamma_t \right)^2 \frac{\left( C_{1, \chi \NN}^{(\text{SI}, s)} + C_{2, \chi \NN}^{(\text{SI}, s)} \right)}{C_{1, \chi N}^{(PT)}} \\
        F_{\text{SD}}(q^2) & = \frac{4 \gamma_t}{q} \tan^{-1} \left( \frac{q}{4 \gamma_t} \right) - \frac{m_N \gamma_t}{2 \pi} C_2^{(\threeS)} \left( \mu - \gamma_t \right)^2 \left[ 1 - \frac{4 \gamma_t}{q} \tan^{-1} \left( \frac{q}{4 \gamma_t} \right) \right] + \frac{\gamma_t}{\pi} \left( \mu - \gamma_t \right)^2 \frac{C_{1, \chi NN}^{\left( \text{SD}, \, s \right)}}{C_{3, \chi N}^{(PT)} } \, .
    \end{align}

In Eq.~\eqref{eq:cross_section:deuteron_SI} both terms are $O(\Nc)$, but the second term is higher order in \eftnopi.
In Eq.~\eqref{eq:cross_section:deuteron_SD} both terms are $O(1)$, and the second term is again higher order in \eftnopi.
Therefore, the ratio of the cross sections is on the order of
    \begin{equation}
            \label{eq:cross_section:deuteron:SD/SI}
        \frac{\sigma^{\text{SD}}_{0,d}}{\sigma^{\text{SI}}_{0,d}} \sim \frac{2}{\Nc^2} \, ,
    \end{equation}
and this is expected to receive $O(1/\Nc^2)$ corrections as well as \eftnopi corrections.
Additionally, the ratio of the SI deuteron-WIMP cross section to the corresponding nucleon-WIMP cross section at LO in the combined EFT and large-\Nc expansion is
    \begin{equation}
        \frac{\sigma^{\text{SI}}_{0,d}}{\sigma^{\text{SI}}_{0,N}} = \frac{4 m_{\chi d}^2}{m_{\chi N}^2} = \frac{4 m_d^2}{m_N^2} \left( \frac{m_N + m_\chi}{m_d + m_\chi} \right)^2 \, .
    \end{equation}
Thus, the deuteron-WIMP SI cross section is fixed with respect to \Nc, but it is roughly an order of magnitude greater than the nucleon-WIMP SI cross section due to the size of the deuteron mass compared to the nucleon mass.
The ratio of the SD cross sections is
    \begin{equation}
        \frac{\sigma^{\text{SD}}_{0,d}}{\sigma^{\text{SD}}_{0,N}} = \frac{8 m_{\chi d}^2}{3 m_{\chi N}^2} \frac{1}{\left( 1 \pm \frac{C^{(PT)}_{2, \chi N}}{C^{(PT)}_{3, \chi N}} \right)^2} \, .
    \end{equation}
Therefore, the SD cross section for the deuteron is $O(1/\Nc^2)$ suppressed relative to the corresponding nucleon cross section.
However, the combined size of the other factors may obscure this relationship, so this constraint should not be over interpreted.
Furthermore, these results can change depending on the choice of the underlying model at the quark level.
For instance, if the dark matter only interacts with the quark axial current, then this will generate SD and not SI scattering, which clearly contradicts Eq.~\eqref{eq:cross_section:deuteron:SD/SI}.
However, these results may be seen as statements regarding hierarchies or trends in a general scenario.

Until now, we have not focused on a specific underlying interaction at the quark level.
In order to compare to similar analyses from ChEFT \cite{korber_first-principle_2017, Andreoli:2018etf}, we will choose the scalar quark current as a benchmark.
There are no isovector contributions in elastic scattering off of the the deuteron;
however, we will include the isovector contributions to the SI response function when we consider ${}^3$H and ${}^3$He.

For this model, the coupling $C_{1, \chi N}^{(PT)}$ can be matched to the pion-nucleon sigma term, $
\sigma_{\pi N}$.
We use the recent lattice result $\sigma_{\pi N} = 59.6(7.4)$ MeV \cite{Gupta:2021ahb}, which is similar to the value used in Refs.~\cite{korber_first-principle_2017, Andreoli:2018etf}.
In our approach, the affect of this choice will be to alter the relative value of the two-body couplings.
On the other hand, the isovector coupling $C_{4, 
\chi N}^{(PT)}$ can be fixed to the neutron-proton strong mass splitting $\delta m_N = 2.32 \pm 0.17$ MeV \cite{Brantley:2016our}.
The details of this matching can be found in Appendix \ref{appendix:dleft}.

\begin{center}
\begin{figure}[h!]
    \centering
    \includegraphics[width = 0.7\textwidth]{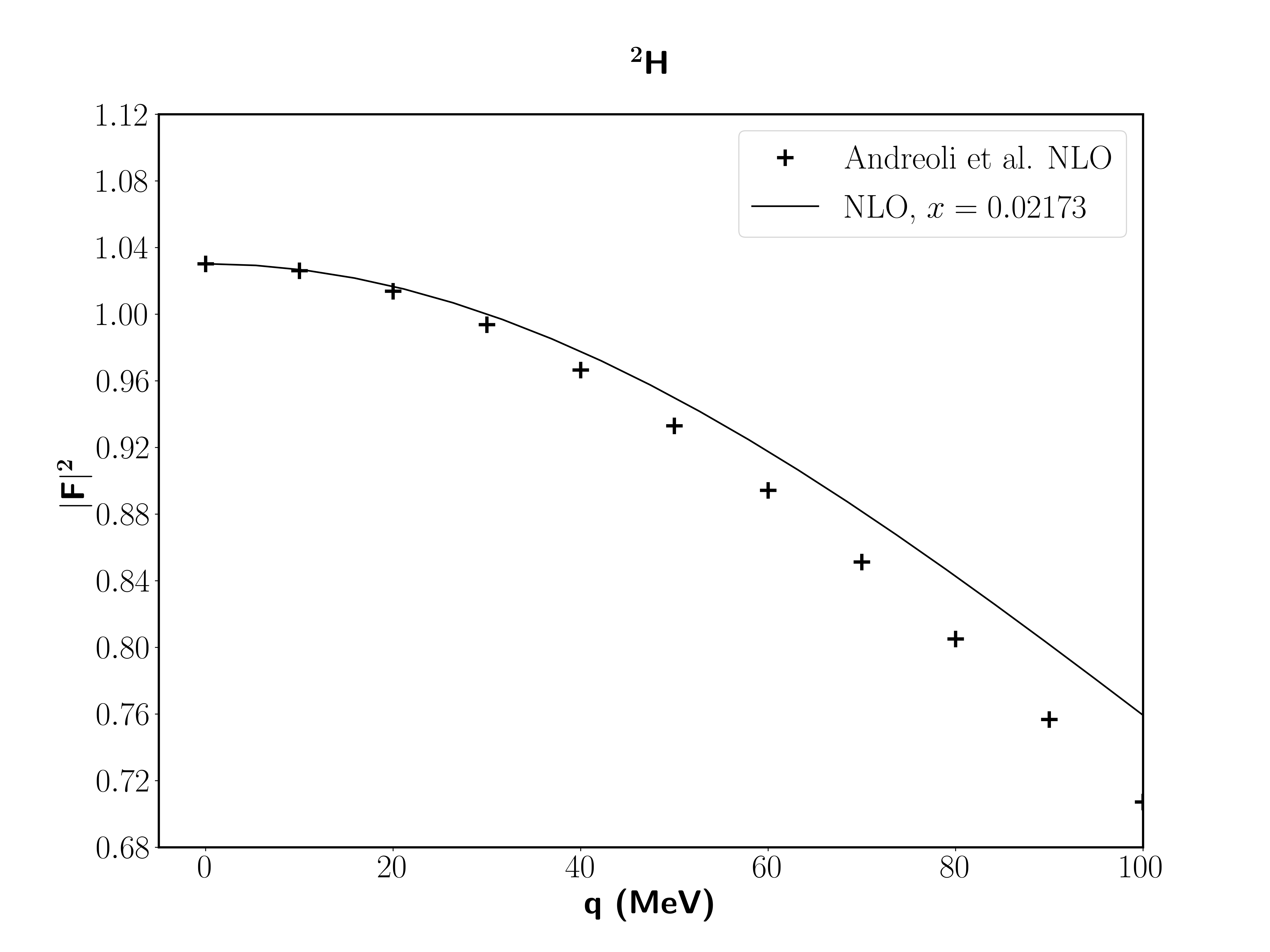}
    \caption{The SI isoscalar response function for the deuteron using the scalar-isoscalar interaction in Refs.~\cite{korber_first-principle_2017, Andreoli:2018etf}. The crosses are taken from Ref.~\cite{Andreoli:2018etf}, the solid line is the \eftnopi result with $x$ fixed to the $q^2 \to 0$ limit of ChEFT data.}
    \label{fig:deuteron_SI_response}
\end{figure}
\end{center}

First, we fix the dimensionless RG invariant combination 
    \begin{equation}
            \label{eq:cross_sections:deuteron:x}
        x = \frac{2 \left( C_{1, \chi \NN}^{(\text{SI}, s)} + C_{2, \chi \NN}^{(\text{SI}, s)} \right)}{m_N C_2^{(\threeS)} C_{1, \chi N}^{(PT)}}
    \end{equation} 
to the $q^2 \to 0$ limit of the corresponding response function in Ref.~\cite{Andreoli:2018etf} using the $Z$ parameterization \cite{Phillips:1999hh}. This leads to $x \approx 0.02$ shown in Fig.~\ref{fig:deuteron_SI_response}, and the resulting response function is in agreement with Ref.~\cite{Andreoli:2018etf} within the \eftnopi errors for all values of $q$ considered.
With the value of $x$ fixed, the large-\Nc constraints may be used to estimate the remaining LEC combinations that receive contributions from the scalar current.
We will include these estimates in the response functions of the next subsection.

%% file: calculations/triton_amplitude.tex
\subsection{Dark matter-${}^3$H and ${}^3$He elastic scattering}
    \label{sec:triton}
    
\begin{figure}[h!]
    \centering
    \includegraphics[width=\textwidth]{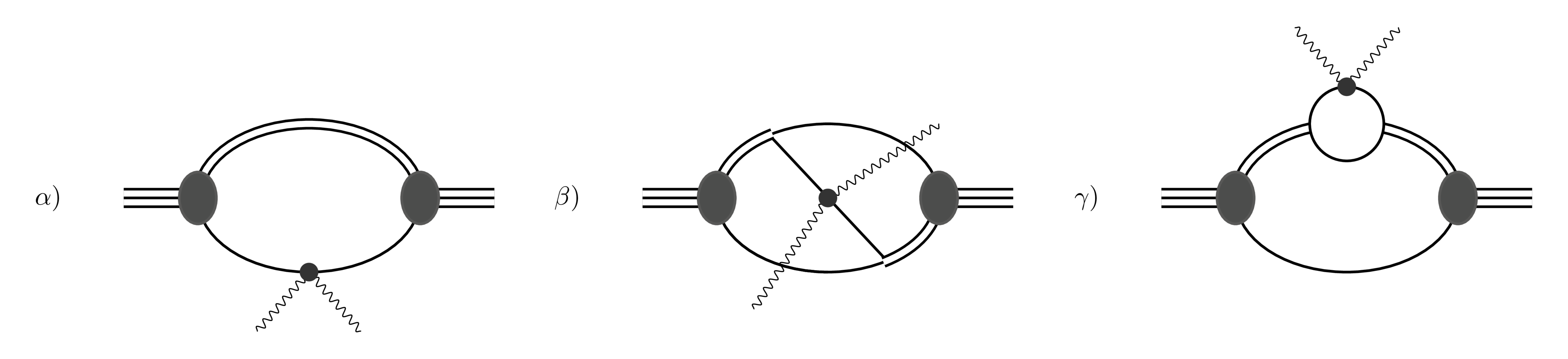}
    \caption{Diagrams that contribute to the WIMP-$^3$H and $^3$He elastic scatterings via the dark-matter-one-nucleon interaction. The double line denotes the dressed propagator of either spin-triplet or spin-singlet dibaryon fields. The gray ovals represent the ${}^3$H/${}^3$He vertex function.}
    \label{fig:dark_matter_triton}
\end{figure}

Finally, we consider elastic WIMP-${}^3$H and WIMP-${}^3$He scattering.
The diagrams are shown in Fig.~\ref{fig:dark_matter_triton}.
In order to perform these calculations, we use the Lagrangian
    \begin{equation}
        \calL = \calL_{\text{strong}} + \calL_{\chi N} + \calL_{\chi ts} \, ,
    \end{equation}
where
\begin{equation}
\begin{aligned}
        \calL_{\text{strong}}  &= \Ndag \left(  i \partial_0 +\frac{\grad^2}{2 m_N} \right) N + t^\dagger_i \left[\Delta_t -  c_{0t} \left( i\partial_0 +\frac{\grad^2}{4 m_N} + \frac{\gamma_t^2}{m_N} \right)\right] t_i + y_t \left[ t_i^\dagger N^TP_iN + \text{H.c} \right]\\
        & + s^\dagger _a\left[\Delta_s -c_{0s} \left( i\partial_0 +\frac{\grad^2}{4 m_N} + \frac{\gamma_s^2}{m_N}\right) \right]s_a  + y_s \left[ s_a^\dagger N^TP_aN + \text{H.c} \right] \\
        & + \psi^\dagger \Omega \psi + \left[ \omega_t \psi^\dagger \sigma_i N t_i - \omega_s \psi^\dagger \tau_a N s_a\ + \text{H.c.} \right] \, ,
    \end{aligned}   
\end{equation}
where $t_i$ is the deuteron field, $s_a$ is the $\oneS$ virtual state dibaryon field, and $\psi$ is an auxiliary spin and isospin doublet field for the three-body bound states. 
In the $Z$ parameterization, the dibaryon-\NN coupling constants are set to $y_s^2=y_t^2=\frac{4\pi}{m_N}$ and the remaining LECs are \cite{griesshammer_improved_2004}
    \begin{align}
        &\Delta_t + \mu = \gamma_t \sim Q \qquad c^{(n)}_{0t} = (-1)^n(Z_t-1)^{n+1}\frac{m_N}{\gamma_t}\sim Q^n\,,\\
        &\Delta_s + \mu = \gamma_s \sim Q \qquad c^{(n)}_{0s} = (-1)^n(Z_s-1)^{n+1}\frac{m_N}{\gamma_s} \sim Q^n\,,
    \end{align}
where $\gamma_t = 45.7025$ MeV is the binding momentum of deuteron, $\gamma_s=-7.8902$ MeV is the binding momentum of the virtual state in the $\oneS$ channel, $Z_t = 1.6908$ and $Z_s=0.9015$ are the residues about the deuteron and $\oneS$ poles, respectively, and $c_{0t/s} = \sum_n c_{0t/s}^{(n)}$.
\begin{figure}[h!]
    \centering
    \includegraphics[width=0.3\textwidth]{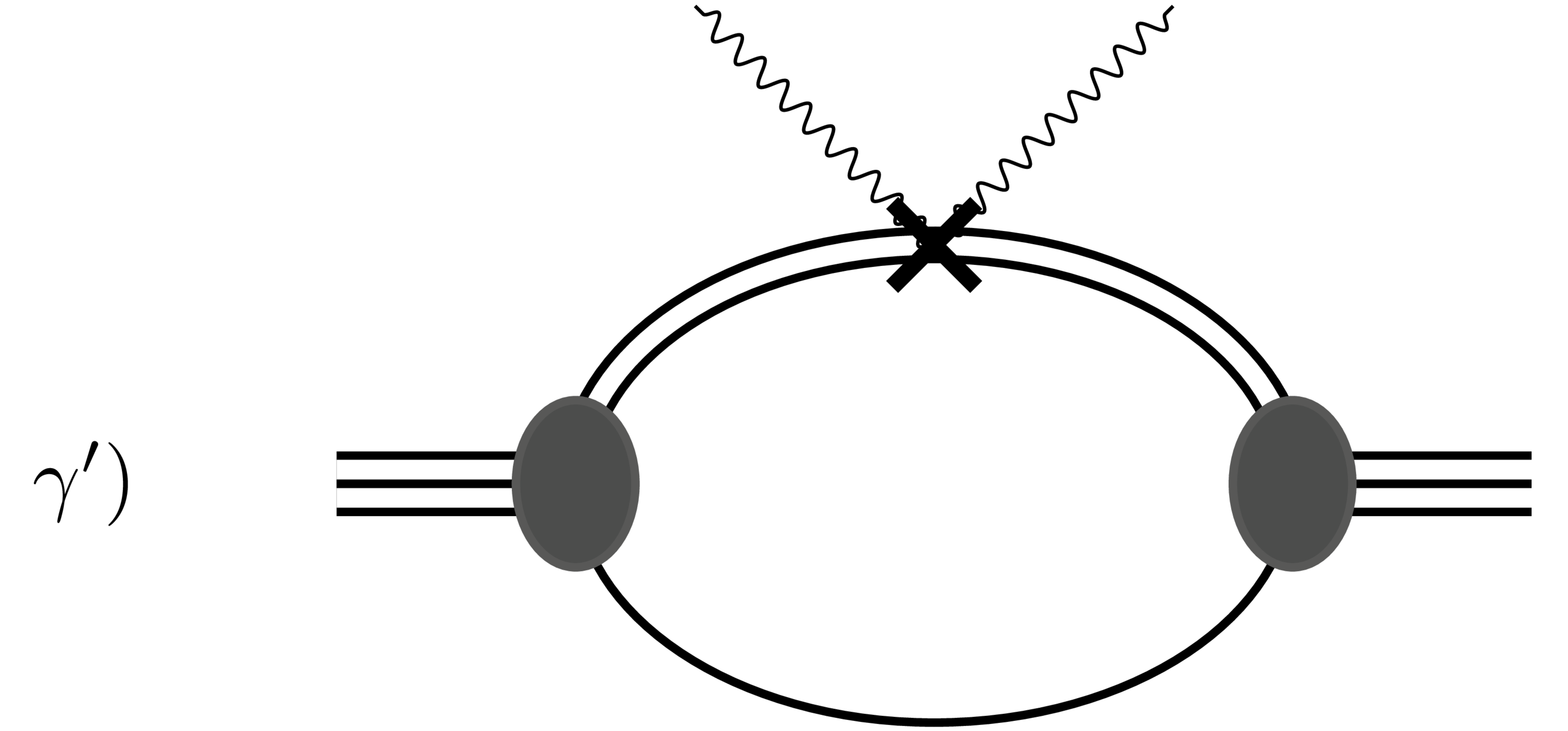}
    \caption{Diagram that contributes at NLO with dark-matter-two-nucleon interaction given by Eq.~\eqref{eq:triton:dibaryon_lagrangian}.}
    \label{fig:dark_matter_triton_2bodycurrent}
\end{figure}
The three-body LECs $\omega_s$ and $\omega_t$ are fixed in order to reproduce the triton binding energy $B_{{}^3\text{H}} = -8.48$ MeV \cite{griesshammer_improved_2004}.
The term $\calL_{\chi N}$ consists of the one-nucleon currents, and $\calL_{\chi NN}$ consists of the two-body currents which takes the form
    \begin{equation}
            \label{eq:triton:dibaryon_lagrangian}
        \begin{aligned}
            \calL_{\chi ts} & = l^\chi_1  \chi^\dagger \chi  t_i^\dagger t_i + il^\chi_2\epsilon_{ijk} \left( \chi^\dagger \sigma^i \chi \right) t_j^\dagger t_k + l^\chi_3 \left( \chi^\dagger \sigma^i \chi \right) \left( s_3^\dagger t_i + \text{H.c.} \right) + l^\chi_4 \left( \chi^\dagger \chi \right) s_a^\dagger s_a \\
            & + l^\chi_5 i \epsilon^{3ab}  \left( \chi^\dagger \chi \right) s_a^\dagger s_b + l^\chi_6 \mathcal{I}_{ab}\left( \chi^\dagger \chi \right) s_a^\dagger s_b + l^\chi_7 \left( \chi^\dagger \sigma^i \chi \right) \left( i s_3^\dagger t_i + \text{H.c.} \right) ,
        \end{aligned}
    \end{equation}
in the dibaryon formalism. 
The relevant Feynman diagram corresponding to an insertion of the two-body currents in Eq.~\eqref{eq:triton:dibaryon_lagrangian} is shown in Fig.~\ref{fig:dark_matter_triton_2bodycurrent}.
An example of matching the LECs in this formalism to those in Eq.~\eqref{eq:Lagrangian:two_nucleon_partial} which is required to apply the large-\Nc scaling rules is presented in Appendix \ref{appendix:dibaryon}.

The three-body vertex function, which is equivalent to the wavefunction, is expressed as a system of coupled inhomogeneous integral equations in cluster configuration space as \cite{griesshammer_improved_2004} 
    \begin{equation}
        \boldsymbol{\calG}_0(k,p) = \vb{B}_0 + \frac{1}{2\pi^2} \int_0^\Lambda dq q^2~\vb{K}(k,p,q ) \boldsymbol{\calG}_0(k,q) \, ,
            \label{eq:vertex_function}
    \end{equation}
where 
    \begin{align}
        &\boldsymbol{\calG}_0(k,p) = \mqty[\calG_t(k,p)\\\calG_s(k,p)],\qquad  \vb{B}_0 = \mqty[1\\-1] \, , 
    \end{align}
The function $\calG_t(k,p)$ is the vertex function describing the coupling between triton, nucleon, and deuteron while $\calG_s(k,p)$ is the vertex function describing the coupling between triton, nucleon, and spin-singlet dibaryon.
The kernel is
    \begin{align}
        &\vb{K}(k,p,q) = \frac{2\pi}{pq}Q_0\left( \frac{p^2+q^2-m_NE -i\epsilon}{pq}\right)
        \begin{bmatrix}-D_t(E-\frac{\vb{q}^2}{2m_N},\vb{q}) & 3D_t(E-\frac{\vb{q}^2}{2m_N},\vb{q})\\3D_s(E-\frac{\vb{q}^2}{2m_N},\vb{q}) & -D_s(E-\frac{\vb{q}^2}{2m_N},\vb{q})\end{bmatrix} \, ,
    \end{align}
where $D_t$ and $D_s$ are the dressed deuteron and spin-singlet dibaryon propagators, respectively, given by
    \begin{equation}
        iD_{t/s}(p^0,\vb{p}) = \frac{i}{\gamma_{t/s} - \sqrt{\frac{\vb{p}^2}{4}-m_Np_0-i\epsilon}} \, ,
    \end{equation}
and the Legendre function of the second kind is
    \begin{equation}
        Q_0(a) = \frac{1}{2}\ln \left( \frac{a+1}{a-1} \right) \, .
    \end{equation}
Finally, the renormalized LO vertex function is obtained by 
    \begin{eqnarray}
        \boldsymbol{\Gamma}_0(p,E) &=& \boldsymbol{\calG}_0(p,E)\sqrt{Z_\psi^{\text{LO}}} \, , 
    \end{eqnarray}
where $Z_\psi^{\text{LO}}$ is the residue of the dressed triton propagator about the triton pole. 
The couplings $\omega_t$ and $\omega_s$ have been absorbed into $Z_\psi^{\text{LO}}$, so they do not appear explicitly in Eq.~\eqref{eq:vertex_function}. 
The calculation of $Z_\psi^{\text{LO}}$ is shown in Ref.~\cite{vanasse_triton_2017}.  
With these ingredients, the diagrams in Fig.~\ref{fig:dark_matter_triton} can be calculated using the tools established in Refs.~\cite{Hagen:2013xga, vanasse_triton_2017}. 
Greater detail concerning the calculations can be found in the Appendix \ref{appendix:triton}.

In the limit of zero momentum transfer, the ratio of the SI and SD cross sections at LO in the combined large-\Nc and \eftnopi expansion is
    \begin{equation}
        \frac{\sigma^{\text{SI}}_{0,H}}{\sigma^{\text{SD}}_{0,H}} \sim \frac{9C^{(PT)2}_{1, \chi N} }{3 C^{(PT)2}_{2, \chi N} } \sim 3 \, ,
    \end{equation}
which is fixed with respect to \Nc.
Similarly, the ratio of the SI WIMP-${}^3$H/${}^3$He cross section to the WIMP-deuteron SI cross section is
    \begin{equation} 
        \frac{\sigma^{\text{SI}}_{0,H}}{\sigma^{\text{SI}}_{0,d}} = \frac{9 m_{\chi H}^2}{4 m_{\chi d}^2} = \frac{9 m_{H}^2}{4 m_d^2} \left( \frac{m_d + m_\chi}{m_{H} + m_\chi} \right)^2 \, ,
    \end{equation}
which is fixed with respect to \Nc.
Lastly, the ratio of the SD cross sections is
    \begin{equation}
        \frac{\sigma^{\text{SD}}_{0,H}}{\sigma^{\text{SD}}_{0,d}} = \frac{3 m_{\chi H}^2}{8 m_{\chi d}^2} \frac{C^{(PT) 2}_{2, \chi N}}{C^{(PT) 2}_{3, \chi N}} \, .
    \end{equation}
Therefore, the deuteron-WIMP SD cross section is relatively $1/\Nc^2$ suppressed.
These relative scalings are similar to the ratios of the deuteron and nucleon cross sections.

\begin{center}
\begin{figure}
    \centering
    \includegraphics[width=0.8\textwidth]{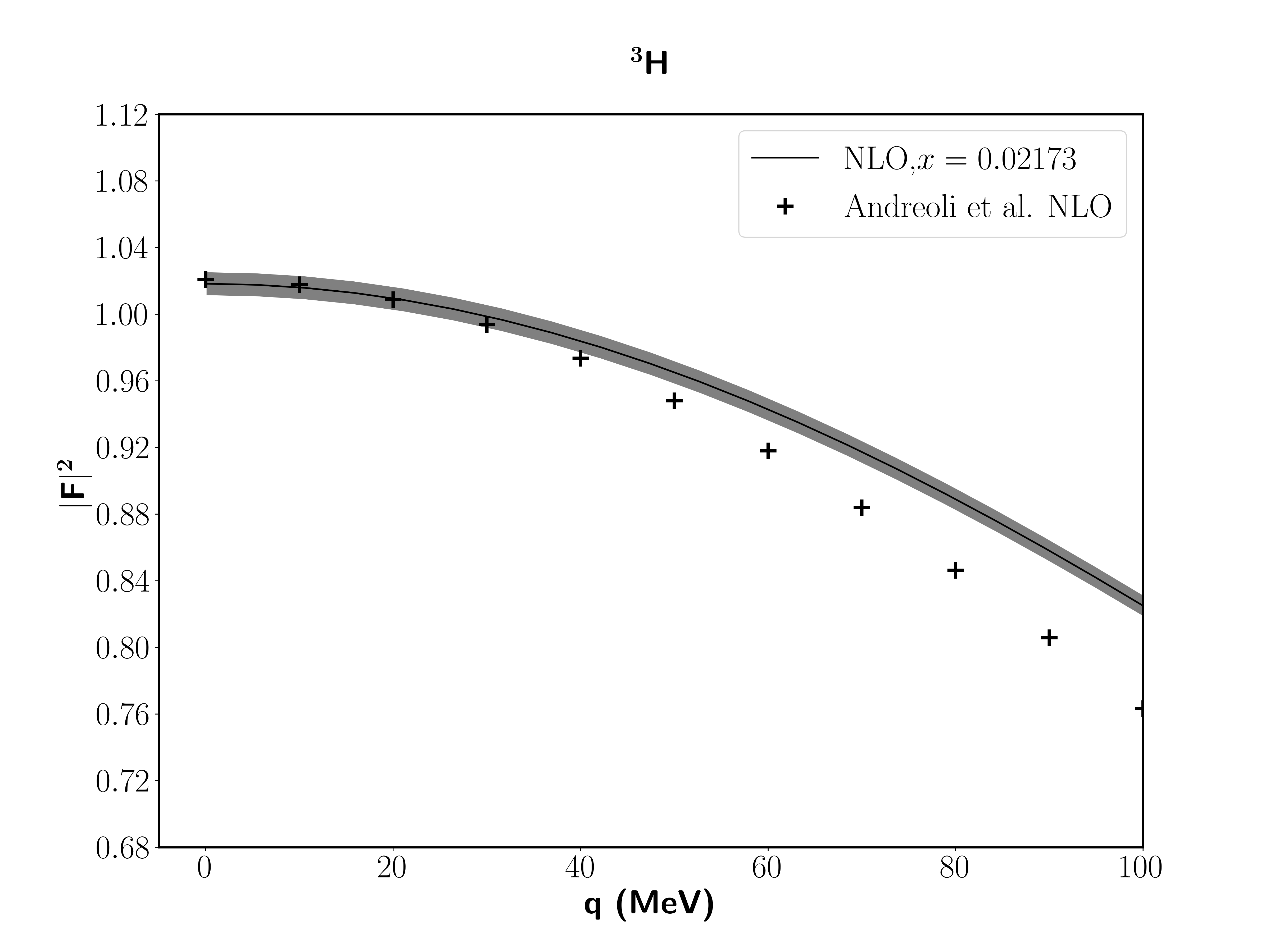}
    \caption{The response function of ${}^3$H for the quark scalar current omitting isovector interactions. The crosses are NLO data from Ref.~\cite{Andreoli:2018etf}. The gray band indicates $30\%$ error from the large-\Nc prediction for $l^\chi_4/l^\chi_1$.}
    \label{fig:3H_response_isoscalar}
\end{figure}
\end{center}

\begin{center}
\begin{figure}
    \centering
    \includegraphics[width=0.8\textwidth]{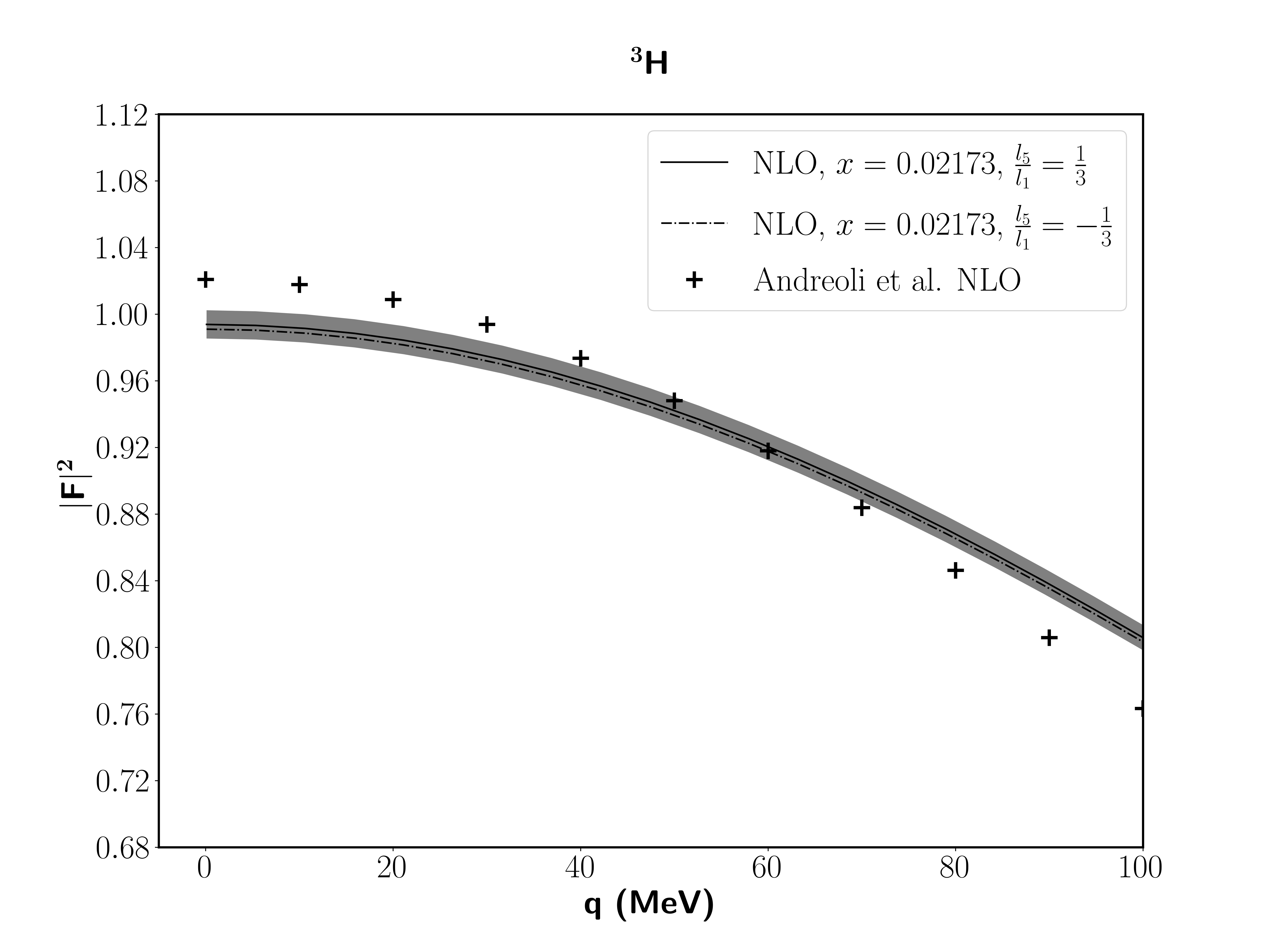}
    \caption{The SI response function of ${}^3$H including isovector contributions. The crosses are the same as Fig.~\ref{fig:deuteron_SI_response}. The solid and dashed lines include the large-\Nc estimate for $l^\chi_5$ with respect to $l^\chi_1$ and with relative positive and negative signs, respectively.}
    \label{fig:3H_response_with_isovector}
\end{figure}
\end{center}

\begin{center}
\begin{figure}
    \centering
    \includegraphics[width=0.8\textwidth]{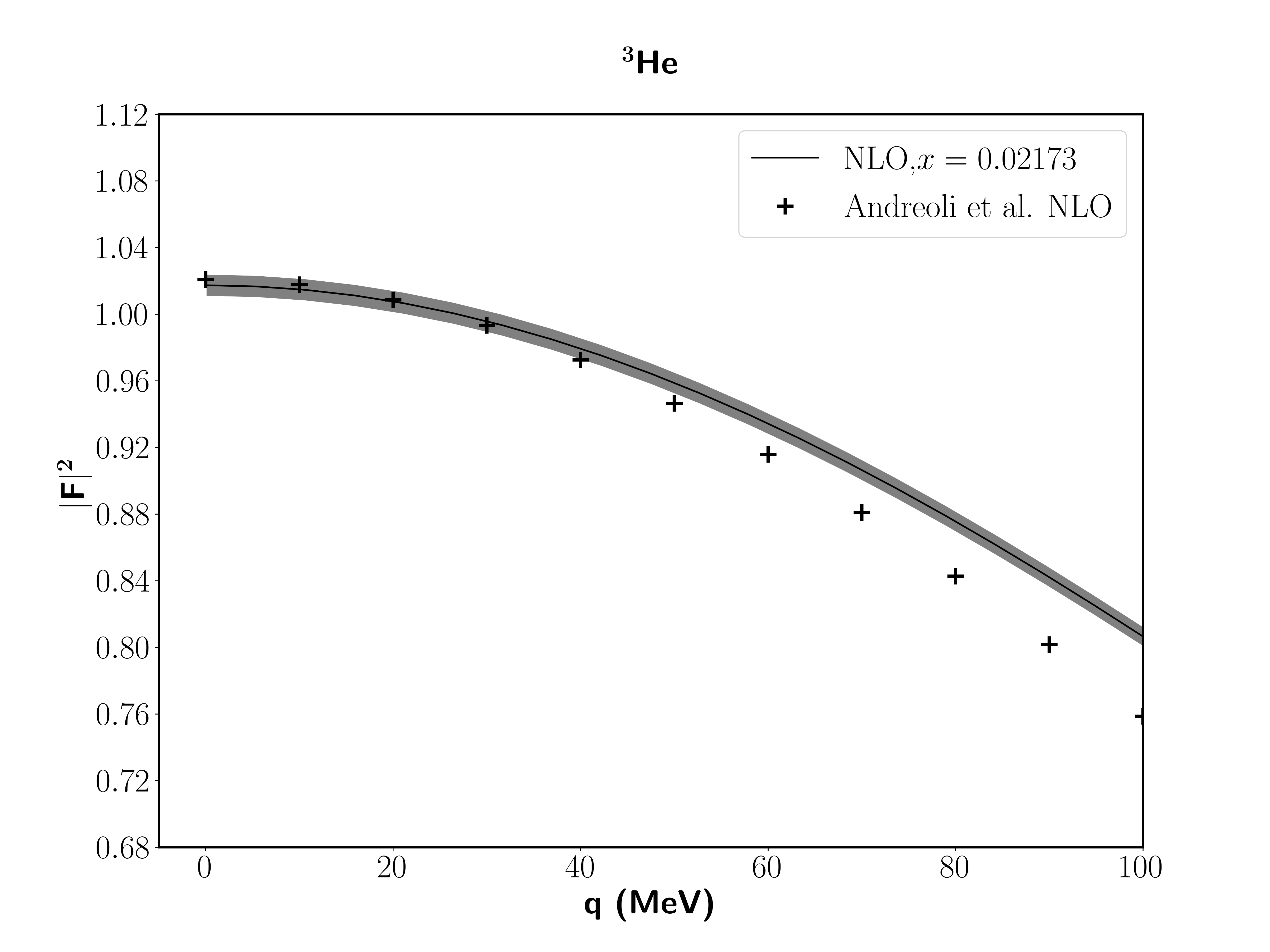}
    \caption{The SI response function of ${}^3$He. The features of the plot are the same as Fig. \ref{fig:3H_response_isoscalar}}. 
    \label{fig:3He_response_isoscalar}
\end{figure}
\end{center}

\begin{center}
\begin{figure}
    \centering
    \includegraphics[width=0.8\textwidth]{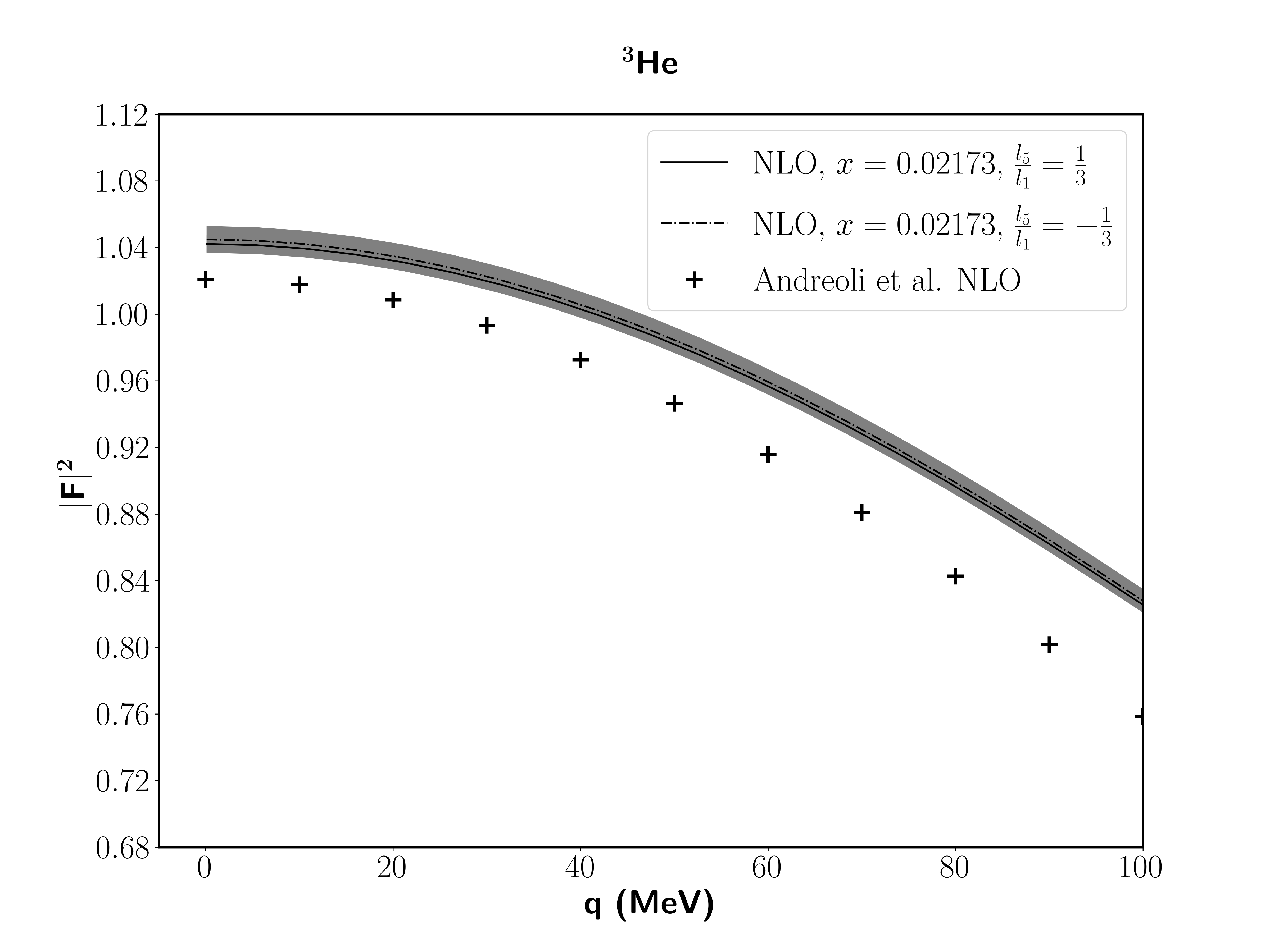}
    \caption{The SI response function for ${}^3$He including isovector contributions. The features of the plot are the same as Fig.~\ref{fig:3H_response_with_isovector}.}
    \label{fig:3He_response_with_isovector}
\end{figure}
\end{center}

The response functions for ${}^3$H and ${}^3$He for a WIMP coupled to the quark scalar isoscalar current are shown in Figs.~\ref{fig:3H_response_isoscalar} and \ref{fig:3He_response_isoscalar}, respectively, where we have used the value of $x$ determined from the deuteron response function in order to fix $l^\chi_1$ and used the large-\Nc estimate $l^\chi_4/l^\chi_1 = 1$.
The result of Ref.~\cite{Andreoli:2018etf} falls within the large-\Nc error band, although this should not be considered as a rigorous uncertainty estimate.
This demonstrates that the large-\Nc estimates can provide reliable results for two-nucleon currents, at least for the two-body contributions to the scalar current.

We have also included the corrections to the SI response functions due to the isovector interaction proportional to $l^\chi_5$ in Figs.~\ref{fig:3H_response_with_isovector} and \ref{fig:3He_response_with_isovector}, where we have used the large-\Nc estimate $\abs{l^\chi_5/l^\chi_1} = 1/3$.
Overall, the inclusion of this operator leads to a small but distinguishable shift in the response functions for the two nuclei.

%% file: conclusion.tex
\section{Conclusions} 
    \label{sec:conclusions}

The detection of dark matter is a high priority in searches for BSM physics.
Direct detection searches have set stringent limits on the WIMP-nucleon interactions subject to strict assumptions regarding the isospin properties of the couplings \cite{PandaX_2017,LUX_2016,Deap_2018,Xenon1T_2017,CDMS_2018, XENON:2019gfn}.
However, new detectors \cite{guo_concept_2013, ito_scintillation_2013, gerbier_NEWS_2014, profumo_GeV_2016, hertel_direct_2019,Maris_Helium_Evaporation_2017} using light nuclei as targets and cosmological studies make it possible to exploit few-body systems for dark matter detections for a wide range of dark matter masses.
Pionless EFT is a model independent framework to study WIMP-light nucleus scattering.
Compared to previous EFT calculations \cite{prezeau_new_2003, cirigliano_wimp-nucleus_2012,Klos_large-scale_2013,hoferichter_chiral_2015, korber_first-principle_2017, hoferichter_nuclear_2019, menendez_spin-dependent_2012, bishara_chiral_2017, bishara_from-quarks_2017, hoferichter_analysis_2016,fan_non-relativistic_2010,fitzpatrick_effective_2013, fitzpatrick_model-independent_2012, anand_model-independent_2015, anand_weakly_2014, hill_standard_2015}, \eftnopi has the advantage that its power counting is well understood and it is renormalization group invariant.
However, all EFT calculations are limited by the lack of a clear signal in experiments that can fix the values of the LECs.
Therefore, theoretical constraints from other means are necessary in order to guide the interpretation of data.

Here, we used the spin-flavor symmetry of the baryon sector of large-\Nc limit of QCD \cite{dashen_1/nc_1994, dashen_baryon_1993, dashen_spin_1995, carone_spin_1994, gervais_large-_1984, gervais_large-_1984-1, luty_baryons_1994} to constrain the relative sizes of the LECs for zero- and one-derivative one-nucleon currents coupled to an external WIMP field. 
If we include 1/\Nc corrections from the SI isovector coupling, the WIMP-neutron and WIMP-protons couplings can move into the xenonphobic regime where potential signals in xenon-based detectors will be highly suppressed.
We also examined the zero-derivative two-nucleon contact currents coupled to a WIMP.
There are three LECs that are $O(\Nc)$ while the remaining LECs are at least relatively $1/\Nc$ suppressed. 
In particular, the SI isoscalar, SD isovector, and SI isotensor LECs are $O(\Nc)$, so isospin violating WIMP-two-nucleon interactions can produce significant corrections to the LO results for light nuclei.

We estimated the impact of the large-\Nc constraints by calculating the elastic WIMP scattering cross sections off the nucleon, deuteron, ${}^3$H, and ${}^3$He.
For the nucleon and the three-body bound states, we found that the SI and SD cross sections are of the same order in \Nc while the SD cross section for the deuteron is $1/\Nc^2$ suppressed relative to the SI cross section.
Moreover, the SI cross sections for each target are of the same size while the deuteron SD cross section is 1/$\Nc^2$ suppressed relative to the WIMP-nucleon and WIMP-${}^3$H/${}^3$He cross sections.
The relative differences in size of the SI cross sections at \LONc is due to the differences in the target masses.
We expect that these results will receive roughly 30$\%$ corrections at NLO in the large-\Nc and \eftnopi expansion.

We have also calculated the response functions for a model in which the quark scalar current couples to the external WIMP field.
This is the same model explored in Refs.~\cite{korber_first-principle_2017, Andreoli:2018etf}, which we used as input data to fix a combination of LECs of our one- and two-body currents relevant for WIMP-deuteron scattering.
The resulting response function for the deuteron agrees with Refs.~\cite{korber_first-principle_2017, Andreoli:2018etf} within \eftnopi errors for a considerable range of momentum transfers, cf. Fig.~\ref{fig:deuteron_SI_response}.
Using the large-\Nc scalings from Eq.~\eqref{eq:Lagrangian:two_nucleon_partial} and Eqs.~\eqref{eq:dibaryon:l1}-\eqref{eq:dibaryon:l7}, we are able to fix the relative sizes of the LECs that contribute to the response functions for ${}^3$H and ${}^3$He.
We find good agreement for the response functions for these nuclei as well in Figs.~\ref{fig:3H_response_isoscalar} and \ref{fig:3He_response_isoscalar}. 
Moreover, we are able to include the isovector contributions to these response functions by making use of the large-\Nc constraints.
We show the impact of these contributions relative to the isoscalar result of Ref.~\cite{Andreoli:2018etf} in Figs.~\ref{fig:3H_response_with_isovector} and \ref{fig:3He_response_with_isovector}.

In summary, the large-\Nc constraints derived in this paper impose an additional hierarchy on top of the \eftnopi power counting for the LECs that couple WIMPs to few-nucleon systems.
Through the use of \eftnopi, we are able to examine the impact of these constraints on the elastic scattering cross sections for several light nuclei.
We hope that these results can be used to inform experimental searches for dark matter that employ light nuclei as well as searches based on cosmology.
In addition, the large-\Nc constraints for contact currents are applicable in ChEFT.
Therefore, they may simplify the required input in many-body calculations that employ ChEFT interactions for direct detection analyses.

%% file: acknowledgments.tex
\begin{acknowledgments}
    We thank Matthias Schindler, Roxanne Springer, and Jared Vanasse for helpful discussions and comments on the manuscript.
    We would also like to thank Christopher K\"orber and Lorenzo Andreoli for making their data accessible to us.
    This material is based upon work supported by the U.S.~Department of Energy, Office of Science, Office of Nuclear Physics,
    under Award Numbers DE-SC0019647 (T.R.R.) and DE-FG02-05ER41368 (T.R.R., X.L., and S.T.N.).
\end{acknowledgments}

%% file: Appendix/appendix_main.tex
\appendix

\include{Appendix/three_body}

\input{Appendix/dleft}

\input{Appendix/matching2}

%% file: Appendix/three_body.tex
\section{Details of the WIMP-Triton/Helium-3 amplitude}
    \label{appendix:triton}

The contribution from each diagram in Fig.~\ref{fig:dark_matter_triton} can be written as
\begin{equation}
    \begin{aligned}
    iA_\alpha & = iC^{(PT)}_{1,\chi N} \delta_{i,i'} \delta_{a,a'} \delta_{m,n}
    \left(I_{\alpha,\, t\overrightarrow{}t} + I_{\alpha,\, s\overrightarrow{}s}\right)  \\
    & + iC^{(PT)}_{2,\chi N} (\sigma_J)_{i,i'} (\tau_3)_{a,a'} (\sigma^J)_{m,n} 
    \left(-\frac{1}{3} I_{\alpha,\, t\overrightarrow{}t} -\frac{1}{3} I_{\alpha,\,s\overrightarrow{}s}\right)  \\
    & + iC^{(PT)}_{3,\chi N} (\sigma_J)_{i,i'} \delta_{a,a'} (\sigma^J)_{m,n}
    \left(-\frac{1}{3}I_{\alpha,\, t\overrightarrow{}t} + I_{\alpha,\, s\overrightarrow{}s}\right)\\
    & + iC^{(PT)}_{4,\chi N} \delta_{i,i'} (\tau_3)_{a,a'} \delta_{m,n} 
    \left(I_{\alpha,\, t\overrightarrow{}t} - \frac{1}{3} I_{\alpha,\, s\overrightarrow{}s}\right)  \, ,
\end{aligned}
\end{equation}

\begin{equation}
    \begin{aligned}
    iA_\beta & = iC^{(PT)}_{1,\chi N} \delta_{i,i'} \delta_{a,a'} \delta_{m,n} 
    \left(-\frac{1}{2}I_{\beta,\, t\overrightarrow{}t} -\frac{1}{2}I_{\beta,\, s\overrightarrow{}s} + \frac{3}{2}I_{\beta,\, t\overrightarrow{}s} + \frac{3}{2}I_{\beta,\, s\overrightarrow{}t}\right)  \\
    & + iC^{(PT)}_{2,\chi N} (\sigma_J)_{i,i'} (\tau_3)_{a,a'} (\sigma^J)_{m,n}
    \left(\frac{5}{6}I_{\beta,\, t\overrightarrow{}t} + \frac{5}{6}I_{\beta,\, s\overrightarrow{}s}+ \frac{1}{6}I_{\beta,\, t\overrightarrow{}s} + \frac{1}{6}I_{\beta,\, s\overrightarrow{}t}\right) \\
    & + iC^{(PT)}_{3,\chi N} (\sigma_J)_{i,i'} \delta_{a,a'} (\sigma^J)_{m,n}
    \left(-\frac{5}{6}I_{\beta,\, t\overrightarrow{}t} +\frac{1}{2}I_{\beta,\, s\overrightarrow{}s} +\frac{1}{2}I_{\beta,\, t\overrightarrow{}s} +\frac{1}{2}I_{\beta,\, s\overrightarrow{}t}\right)\\
    & + iC^{(PT)}_{4,\chi N} \delta_{i,i'} (\tau_3)_{a,a'} \delta_{m,n} 
    \left(\frac{1}{2}I_{\beta,\, t\overrightarrow{}t} -\frac{5}{6}I_{\beta,\, s\overrightarrow{}s} + \frac{1}{2}I_{\beta,\, t\overrightarrow{}s} + \frac{1}{2}I_{\beta,\, s\overrightarrow{}t}\right)  \, ,
\end{aligned}
\end{equation}
and
\begin{equation}
    \begin{aligned}
     iA_\gamma & = iC^{(PT)}_{1,\chi N} \delta_{i,i'} \delta_{a,a'} \delta_{m,n} 
    (2I_{\gamma,\, t\overrightarrow{}t} + 2I_{\gamma,\, s\overrightarrow{}s})  \\
    & + iC^{(PT)}_{2,\chi N} (\sigma_J)_{i,i'} (\tau_3)_{a,a'} (\sigma^J)_{m,n} 
    \left(\frac{2}{3} I_{\gamma,\, t\overrightarrow{}s} +\frac{2}{3} I_{\gamma,\,s\overrightarrow{}t}\right)   \\
    & + iC^{(PT)}_{3,\chi N} (\sigma_K)_{i,i'} \delta_{a,a'}(\sigma^K)_{m,n}
   \left (\frac{4}{3} I_{\gamma,\, t\overrightarrow{}t}\right) \\
    & + iC^{(PT)}_{4,\chi N} \delta_{i,i'} (\tau_3)_{a,a'} \delta_{m,n} 
    \left( \frac{4}{3} I_{\gamma,\, s\overrightarrow{}s}\right), 
\end{aligned}
\end{equation}
where $I_{\nu, \, x \to y}$ is the integral from diagram $\nu$ and the first dibaryon is of type $x$ while the second dibaryon is of type $y$. 
Then we define $\calI_i$ to be the sum of contributions proportional $C^{(PT)}_{i,\chi N}$. 
For example, $\calI_1$ is given as,
\begin{equation}
    \begin{aligned}
    \calI_1  &= I_{\alpha,\,t\rightarrow t} + I_{\alpha,\, s\rightarrow s}  \\ 
    & -\frac{1}{2}I_{\beta,\, t\overrightarrow{}t} -\frac{1}{2}I_{\beta,\, s\overrightarrow{}s} + \frac{3}{2}I_{\beta,\, t\overrightarrow{}s} + \frac{3}{2}I_{\beta,\, s\overrightarrow{}t}\\
    & + 2I_{\gamma,\, t\overrightarrow{}t} + 2I_{\gamma,\, s\overrightarrow{}s}\,.
\end{aligned}
\end{equation}
We choose to perform the calculation of the nuclear form factor in the Breit frame. In momentum space, the integral for diagram $\alpha$ in Fig.~\ref{fig:dark_matter_triton}$\alpha$ takes the form
\begin{equation}
  I_\alpha = i\int \frac{d^4l}{(2\pi)^4}  \boldsymbol{\Tilde{\calG}}_0\left[\left(E,-\frac{\vb{q}}{2}\right) ,(l_0,\vb{l}) \right]\Sigma_\alpha 
 \left(E,l_0,\vb{l},\vb{q}\right) \boldsymbol{\Tilde{\calG}}_0\left[\left(E,\frac{\vb{q}}{2}\right), (l_0,\vb{l})\right],
\end{equation}
where $\boldsymbol{\Tilde{\calG}}_0$ is the LO vertex function calculated in an arbitrary frame, $E = E_3 +\frac{q^2}{24m_N}$, and $\Sigma_\alpha \left(E,l_0,\vb{l},\vb{q}\right)$ is given by
 
\begin{align}
\Sigma_\alpha &= \frac{1}{E-l_0 -\frac{(\vb{l}+\vb{q}/2)^2}{2m_N}+i\epsilon}~\frac{1}{\gamma_{t/s}-\sqrt{\frac{l^2}{4}-m_Nl_0+i\epsilon}}~\frac{1}{E-l_0 -\frac{(\vb{l}-\vb{q}/2)^2}{2m_N}+i\epsilon}. 
\end{align}
Here, $l$ and $q$ are the magnitude of the vectors $\vb{l}$ and $\vb{q}$, respectively. We perform the $l_0$ integral over the energy pole, which puts the left-hand nucleon on-shell. Next, 
integration over the azimuthal angle yields
\begin{align}
I_\alpha = \frac{M}{2\pi^2}\int_0^\Lambda d ll^2\int_{-1}^1 d\cos\theta~\boldsymbol{\Tilde{\calG}}_0\left[\left(E,-\frac{\vb{q}}{2}\right),(l_0,\vb{l}) \right]&\frac{1}{\gamma_{t/s}-\sqrt{\frac{3}{4}l^2-m_NE+\frac{1}{8}q^2+\frac{1}{2}lq\cos\theta}}\nonumber\\
&~\times\frac{1}{lq\cos\theta}~\boldsymbol{\Tilde{\calG}}_0\left[\left(E,\frac{\vb{q}}{2}\right),(l_0,\vb{l}) \right],
\end{align}
where $\Lambda$ is a cutoff used to regulate UV divergences.
Finally, Galilean invariance is used to relate the vertex function in an aribtrary frame to the vertex function in the center-of-mass frame, $\boldsymbol{\calG}_0$.
If the nucleon pole is enclosed in the contour of integration for the the vertex function, then
\beq
\boldsymbol{\Tilde{\calG}}_0\left[\left(E,-\frac{\vb{q}}{2}\right),(l_0,\vb{l}) \right] = \boldsymbol{\calG}_0\left(\left|\vb{l} + \frac{\vb{q}}{3}\right|\right) \, ,
\eeq
otherwise, 
\beq
\boldsymbol{\Tilde{\calG}}\left[\left(E,\frac{\vb{q}}{2}\right),(l_0,\vb{l}) \right] = \boldsymbol{\calG}_0\left(E_3 -\frac{q^2}{18m_N}-\frac{2}{3m_N}lq\cos\theta-\frac{l^2}{2m_N},\left|\vb{l} - \frac{\vb{q}}{3}\right|\right).
\eeq

The integral for diagram $\beta$ has the form
\begin{align}
    I_\beta  = \int \frac{\dd[4]{l}}{(2\pi)^4} & \int \frac{\dd[4]{k}}{(2\pi)^4}  \boldsymbol{\Tilde{\calG}}_0\left[\left(E,-\frac{\vb{q}}{2}\right),(l_0,\vb{l}) \right] ~\Sigma_\beta 
 \left(E,l_0,k_0,\vb{l},\vb{k},\vb{q}\right) ~\boldsymbol{\Tilde{\calG}}_0\left[\left(E,\frac{\vb{q}}{2}\right),(k_0,\vb{k})\right],
    \end{align}
where 
\begin{align}
 \Sigma_\beta = D_N&\left(E-l_0,-\vb{l}-\frac{\vb{q}}{2}\right)~D_{t/s}(l_0,\vb{l}) ~ D_N\left(E-l_0 +k_0,\vb{l}-\frac{\vb{q}}{2} +\vb{k}\right)\nonumber\\
    & \times D_{t/s}(k_0,\vb{k}) ~  D_N\left(E-k_0,-\vb{k}+\frac{\vb{q}}{2}\right) D_N\left(-E+l_0+k_0,\vb{l}+\frac{\vb{q}}{2}+\vb{k}\right) \,.   
\end{align}
Using contour integration over $k_0$ and $l_0$ and enclosing the poles at
\beq
l_0 = E-\frac{(\vb{l}+\vb{q}/2)^2}{2m_N},\quad k_0 = E-\frac{(\vb{k}-\vb{q}/2)^2}{2m_N} \, ,
\eeq
yields
\begin{align}
    I_\beta  = \int \frac{\dd[3]{\vb{l}}}{(2\pi)^3} & \int \frac{\dd[3]{\vb{k}}}{(2\pi)^3}~ \boldsymbol{\Tilde{\calG}}_0\left[\left(E,-\frac{\vb{q}}{2}\right),(l_0,\vb{l}) \right] ~ D_{t/s}(l_0,\vb{l}) ~ D_N\left(E-l_0 +k_0,\vb{l}-\frac{\vb{q}}{2} +\vb{k}\right)\nonumber\\
    & \times D_{t/s}(k_0,\vb{k}) ~  D_N\left(-E+l_0+k_0,\vb{l}+\frac{\vb{q}}{2}+\vb{k}\right) ~\boldsymbol{\Tilde{\calG}}_0\left[\left(E,\frac{\vb{q}}{2}\right),(k_0,\vb{k})\right]\,.
    \label{eq:diagramc}
    \end{align}
The vertex functions are again related to the center-of-mass frame according to
\beq
\boldsymbol{\Tilde{\calG}}_0\left[\left(E,-\frac{\vb{q}}{2}\right),(l_0,\vb{l}) \right] = \boldsymbol{\calG}_0\left(\left|\vb{l} + \frac{\vb{q}}{3}\right|\right),
\eeq
\beq
\boldsymbol{\Tilde{\calG}}_0\left[\left(E,\frac{\vb{q}}{2}\right),(k_0,\vb{k})\right] = \boldsymbol{\calG}_0\left(\left|\vb{k} - \frac{\vb{q}}{3}\right|\right),
\eeq
Consider $\vb{k} = \vb{k}_{\perp} +\vb{k}_{\|}$ and $\vb{l} = \vb{l}_{\perp} +\vb{l}_{\|}$, where $\perp$ and $\|$ respectively indicate vectors perpendicular and parallel to the vector $\vb{q}$,
\begin{align}
\vb{k}\cdot\vb{l} =  \vb{k}_\|\cdot\vb{l}_\| +\vb{k}_\perp\cdot\vb{l}_\perp  = kl\cos\theta\cos\psi + kl\sin\theta\sin\psi \cos\phi,
\end{align}
where $\vb{q}\cdot \vb{l} = ql\cos\theta$, $\vb{q}\cdot \vb{k} = qk\cos\psi$, and $\phi$ is the angle between $\vb{k}_\perp$ and $\vb{l}_\perp$. We will integrate Eq.~\eqref{eq:diagramc} over one azimuthal angle to obtain
\begin{align}
    I_\beta  &= \frac{4m_N^2}{(2\pi)^5}\int_{0}^\Lambda l^2dl \int_0^\Lambda k^2dk \int_{-1}^{1}du \int_{-1}^{1}dv\int_{0}^{2\pi}d\phi ~ \boldsymbol{\calG}_0\left(\left|\vb{l} + \frac{\vb{q}}{3}\right|\right) ~\frac{1}{\gamma_{t/s}-\sqrt{\frac{3}{4}l^2 +\frac{1}{2}lqu +\frac{1}{8}q^2-m_NE}} \nonumber\\
    &\times \frac{1}{\gamma_{t/s}-\sqrt{\frac{3}{4}k^2 -\frac{1}{2}kqv +\frac{1}{8}q^2-m_NE}} \\
    & \times ~\frac{1}{ 2m_NE- (2k^2 + \frac{1}{4}q^2 -\frac{3}{2}lqu - \frac{3}{2}kqv +2 kluv +2 kl\sqrt{1-u^2}\sqrt{1-v^2}\cos\phi)}\nonumber\\
    & \times \frac{1}{ 2m_NE - (2l^2 + 2k^2 + \frac{3}{4}q^2 +\frac{3}{2}lqu - \frac{1}{2}kqv +2 kluv +2 kl\sqrt{1-u^2}\sqrt{1-v^2}\cos\phi)} \boldsymbol{\calG}_0\left(\left|\vb{k} - \frac{\vb{q}}{3}\right|\right)\nonumber\,.
    \end{align}
In the above equation, we have denoted $u \equiv \cos\theta$ and $v\equiv \cos\psi$ for short. 

Lastly, the integral for diagram $\gamma$ is
\begin{align}
    I_\gamma & = -\int \frac{\dd[4]{l}}{(2\pi)^4}\int\frac{\dd[4]{k}}{{(2\pi)^4}} \boldsymbol{\Tilde{\calG}}_0\left[\left(E,-\frac{\vb{q}}{2}\right), (l_0,\vb{l}) \right]\Sigma_\gamma(E,l_0,k_0,\vb{l},\vb{k},\vb{q})  \boldsymbol{\Tilde{\calG}}_0\left[\left( E,\frac{\vb{q}}{2}\right),(l_0,\vb{q}+\vb{l}) \right],
    \label{eq:integral_diagram_gamma}
\end{align}
where
\begin{equation}
\begin{aligned}
 \Sigma_\gamma = D_N&\left(E-l_0, -\frac{\vb{q}}{2}-\vb{l}\right) D_{t/s}\left(l_0,\vb{l}\right)~ D_N(k_0,\vb{k})~D_N(l_0-k,\vb{l}-\vb{k})\\
 &\times D_N\left(l_0-k_0,\vb{q}+\vb{l}-\vb{k}\right) D_{t/s}\left(l_0,\vb{q}+\vb{l}\right)  \, .
\end{aligned}
\end{equation}
Following the same procedure, we enclose the poles at
\beq
l_0 = E-\frac{(\vb{l}+\vb{q}/2)^2}{2m_N},\quad k_0 =\frac{\vb{k}^2}{2m_N}.
\eeq
The vertex functions are related to the center-of-mass frame,
\begin{align}
\boldsymbol{\Tilde{\calG}}_0\left[\left(E,-\frac{\vb{q}}{2}\right),(l_0,\vb{l}) \right] &= \boldsymbol{\calG}_0\left(\left|\vb{l} + \frac{\vb{q}}{3}\right|\right),    \\
\boldsymbol{\Tilde{\calG}}_0\left[\left(E,\frac{\vb{q}}{2}\right),(l_0,\vb{l}) \right] &= \boldsymbol{\calG}_0\left(\left|\vb{l} - \frac{\vb{q}}{3}\right|\right).  
\end{align}
The nucleon sub-loop integral becomes
\beq
\begin{aligned}
I_\text{sub-loop} &= \int \frac{d^3\vb{k}}{(2\pi)^3}\frac{1}{E-\frac{(\vb{l}+\vb{q}/2)^2}{2m_N}-\frac{ \vb{k}^2}{2m_N} - \frac{(\vb{l}-\vb{k})^2}{2m_N}} \frac{1}{E -\frac{\vb{k}^2}{2m_N}-\frac{(\vb{l}+\vb{q}/2)^2}{2m_N} - \frac{(\vb{q}+\vb{l}-\vb{k})^2}{2m_N} }\\
&=-\frac{m_N^2}{2\pi q}\tan^{-1} \left[\frac{q}{2(a+b)}\right], 
\end{aligned}
\eeq
where 

\begin{align}
a^2 &= \frac{3l^2}{4}+\frac{1}{2}\vb{l}\cdot \vb{q}+\frac{q^2}{8}-m_NE,\\
b^2 &= \frac{3l^2}{4}+\vb{l}\cdot \vb{q}+\frac{3q^2}{8}-m_NE.
\end{align}
Finally, the integral in Eq.~\eqref{eq:integral_diagram_gamma} simplifies to
\begin{align}
   I_\gamma = \frac{m_N^2}{8\pi^3 q} \int_{-1}^1 d(\cos\theta) \int_0^\Lambda dl ~l^2~ \boldsymbol{\calG}_0\left(\left|\vb{l} + \frac{\vb{q}}{3}\right|\right) ~ \boldsymbol{\calG}_0\left(\left|\vb{l} - \frac{\vb{q}}{3}\right|\right)\nonumber\\
   \times \tan^{-1}\left[\frac{q}{2(a+b)}\right]\frac{1}{\gamma_{t/s} - \sqrt{a^2-i\epsilon}}\frac{1}{\gamma_{t/s} - \sqrt{b^2-i\epsilon}},
\end{align}
where $\theta$ is the angle between $\vb{l}$ and $\vb{q}$.

%% file: Appendix/dleft.tex
\section{Matching one-nucleon currents to single-quark currents}
    \label{appendix:dleft}

Here, we collect expressions relevant for connecting the one-nucleon LECs to the low energy dark matter-quark couplings in Ref.~\cite{Brod:2017bsw, bishara_from-quarks_2017, hill_standard_2015} (see also Ref.~\cite{del_nobile_theory_2022}).
The interaction Lagrangian below the electroweak scale is
    \begin{align}
            \label{eq:dim6}
        \calL_{\chi q} & =  
        \sum_{f=u,d} 
        \left( \bar \chi \chi \right) \left[ m_f \left( \bar f C^{(f)}_{SS} f \right) + m_f \left( \bar f C^{(f)}_{SP} i \gamma^5 f \right) \right] + \left( \bar \chi i \gamma^5 \chi \right) \left[ m_f \left( \bar f C^{(f)}_{PS} f \right)  + m_f \left( \bar f i \gamma^5 C^{(f)}_{PP} f \right)  \right] \nonumber \\
        &   + \left( \bar \chi \gamma^\mu \chi \right) \left[ \left( \bar f \gamma_\mu C^{(f)}_{VV} f \right) + \left( \bar f \gamma_\mu \gamma^5 C^{(f)}_{VA} f \right) \right] + \left( \bar \chi \gamma^\mu \gamma^5 \chi \right) \left[ \left( \bar f \gamma_\mu C^{(f)}_{AV} f \right) + \left( \bar f \gamma_\mu \gamma^5 C^{(f)}_{AA} f \right) \right] \nonumber \\
        & + \left( \bar \chi \sigma^{\mu \nu} \chi \right) \left( \bar f C^{(f)}_{TT} \sigma^{\mu \nu} f \right) + \left( \bar \chi \sigma^{\mu \nu} \gamma^5 \chi \right) \left( \bar f C^{(f)'}_{TT} \sigma^{\mu \nu} f \right) ,
    \end{align}
where $m_f$ is the mass of the quark of flavor $f$.
From the point of view of the large-\Nc expansion, contributions from strange quarks and heavy quarks are highly suppressed in processes involving only nucleons.
Therefore, we only consider couplings to up and down quarks in this work.
The single-nucleon matrix elements of the quark currents are
    \begin{align}
        \bra{N'} m_f \bar f f \ket{N} & = m_N F^{(f, N)}_S(q^2) \bar N' N \, , \label{eq:form_factor:scalar} \\
        \bra{N'} m_f \bar f i \gamma^5 f \ket{N} & = m_N F^{(f, N)}_P(q^2) \bar N' i \gamma^5 N \, , \label{eq:form_factor:pseudoscalar} \\
        \bra{N'} \bar f \gamma^\mu f \ket{N} & = \bar N' \left[ F^{(f, N)}_1(q^2) \gamma^\mu + \frac{i}{2m_N} F^{(f, N)}_2(q^2) \sigma^{\mu \nu} q_\nu \right] N \, , \label{eq:form_factor:vector} \\
        \bra{N'} \bar f \gamma^\mu \gamma^5 f \ket{N} & = \bar N' \left[ F^{(f, N)}_A(q^2) \gamma^\mu \gamma^5 + \frac{1}{2m_N} F^{(f, N)}_{P'}(q^2) \gamma^5 q^\mu \right] N \, , \label{eq:form_factor:axial} \\
        \bra{N'} \bar f \sigma^{\mu \nu} f \ket{N} & = \bar N \left[ F_{T,0}^{(f, N)}(q^2) \sigma^{\mu \nu} + \frac{i}{2m_N} F_{T,2}^{(f, N)}(q^2) \gamma^{\left[ \mu \right.} q^{\left. \nu \right]} + \frac{i}{m_N^2} F_{T,2}^{(f, N)}(q^2) q^{\left[ \mu \right.} k^{\left. \nu \right]} \right] N \, \label{eq:form_factor:tensor} \, ,
    \end{align}
where we use the notation $N' = N(p')$ and $q = p' - p$.
Here, $N$ also refers to a specific nucleon rather than the isodoublet of the proton and neutron.

In order to match the \eftnopi LEC $C_{1, 
\chi N}^{(PT)}$ to $\sigma_{\pi N}$, it is convenient to rearrange the scalar current according to
    \begin{align}
        \sum_f C_{\Gamma S} m_f \bar f f & = C^{(s)}_{\Gamma S} \bar m \bar q q - C^{(v)}_{\Gamma S} (m_d - m_u) \bar q \tau^3 q \, , \\
        C^{(s)}_{\Gamma S} & = \frac{1}{2} \left[ C^{(u)}_{\Gamma S} \left( 1 - \xi \right) + C^{(d)}_{\Gamma S} \left( 1 + \xi \right)  \right] \, , \\
        C^{(v)}_{\Gamma S} & = \frac{1}{4} \left[ C^{(u)}_{\Gamma S} \left( 1 - \frac{1}{\xi} \right) + C^{(d)}_{\Gamma S} \left( 1 + \frac{1}{\xi} \right) \right] \, ,
    \end{align}
where $\xi = \frac{m_d - m_u}{m_d + m_u} = 0.35 \pm 0.02$ \cite{Aoki:2021kgd}, and $\Gamma$ indicates the relevant dark matter Dirac matrix.
In the text, we consider $C_{\Gamma S}^{(s)} = C_{\Gamma S}^{(v)}$ for simplicity, but this choice can easily be modified in order to study other scenarios.
Now, taking the matrix element at zero momentum transfer yields
    \begin{align}
        \bra{N} \bar m \bar q q \ket{N} & = \sigma_{\pi N} \, , \\
        \bra{N} (m_d - m_u) \bar q \tau^3 q \ket{N} & = \pm \delta{m_N} \, ,
    \end{align}
where the upper (lower) sign corresponds to the proton (neutron).

In order to finish the matching, we perform a nonrelativistic expansion of the nucleon matrix elements through $O(1/m_N)$
    \begin{align}
        \bar N N & = N^\dagger N \, , \\
        \bar N i \gamma^5 N & = \frac{1}{2m_N} \del^i \left( N^\dagger \sigma^i N \right) \, , \\
        \bar N \gamma^0 N & = N^\dagger N \, , \\
        \bar N \gamma^i N & = -\frac{i}{2m_N} N^\dagger \Galilean^i N + \frac{1}{2 m_N} \epsilon^{ijk} \del^j \left( N^\dagger \sigma^k N \right) \, , \\
        \bar N \gamma^0 \gamma^5 N & = -\frac{i}{2m_N} N^\dagger \sigma^i \Galilean^i N \, , \\
        \bar N \gamma^i \gamma^5 N & = N^\dagger \sigma^i N \, , \\
        \bar N \sigma^{0i} N & = \frac{1}{2m_N} \del^i \left( N^\dagger N \right) - \frac{i}{2m_N} \epsilon^{ijk} N^\dagger \sigma^j \Galilean^k N  \, , \\
        \bar N \sigma^{ij} N & = \epsilon^{ijk} N^\dagger \sigma^k N \, , \\
        \bar N i \sigma^{0i} \gamma^5 N & = -N^\dagger \sigma^i N \, , \\
        \bar N i \sigma^{ij} \gamma^5 N & = \frac{1}{2m_N} \epsilon^{ijk} \del^k \left( N^\dagger N \right) + \frac{i}{2m_N} \left[ N^\dagger \sigma^j \Galilean^i N - N^\dagger \sigma^i \Galilean^j N \right]
    \end{align}
This expansion is derived from a heavy particle expansion such as that performed in nonrelativistic QED and heavy quark effective theory (see, e.g., \cite{Caswell:1985ui, manohar_wise_2000}). 
Analagous reductions hold for the dark matter bilinears.

Taking the relevant combinations of proton and neutron matrix elements yields
    \begin{align}
        C_{1, \chi N}^{(P T)} & = C^{(s)}_{SS} \sigma_{\pi N} \, , \\
        C_{4, \chi N}^{(P T)} & = C^{(v)}_{SS} \delta m_N \, ,
    \end{align}
when we restrict ourselves to the scalar current.
When we include the isovector corrections to the three-body response functions, we merely assume $C^{(s)}_{SS}/C^{(v)}_{SS}=1$.
However, one can easily implement other ratios for these couplings.

It is straightforward to use the remaining form factors and nonrelativistic reductions to derive a general set of matching relations.
A similar procedure, just in a different context, has been carried out in Refs.~\cite{hill_standard_2015, bishara_chiral_2017, bishara_from-quarks_2017, Bishara:2018vix, Brod:2017bsw}.
The coefficients of the operators in Eq.~\eqref{eq:dim6} match onto the zero-derivative one-nucleon currents of Eq.~\eqref{eq:Lagrangian:Single_Nucleon:LO} in the following way:
    \begin{align}
        C_{1, \chi N}^{(PT)} & = C_{SS}^{(s)} \sigma_{\pi N} + C_{VV}^{(s)} F_1^{(s)} \, , \\
        C_{2, \chi N}^{(PT)} & = C_{AA}^{(v)} F_A^{(v)} + 2 C_{TT}^{(v)} F_{T,0}^{(v)} \, , \\
        C_{3, \chi N}^{(PT)} & = C_{AA}^{(s)} F_A^{(s)} + 2 C_{TT}^{(s)} F_{T,0}^{(v)} \, , \\
        C_{4, \chi N}^{(PT)} & = C_{SS}^{(v)} \delta m_N + C_{VV}^{(v)} F_1^{(s)} \, .
    \end{align}
The $\slashed P T$ one-nucleon currents receive the contributions
    \begin{align}
        C_{5, \chi N}^{(\slashed P T)} & = - \frac{1}{2m_N} C_{AV}^{(v)} \left( F_1^{(v)} + F_2^{(v)} \right) + \frac{1}{2 m_\chi} C_{VA}^{(v)} F_A^{(v)} \, , \\
        C_{6, \chi N}^{(\slashed P T)} & = - \frac{1}{2m_N} C_{AV}^{(s)} \left( F_1^{(s)} + F_2^{(s)} \right) + \frac{1}{2 m_\chi} C_{VA}^{(s)} F_A^{(s)} \, , \\
        C_{7, \chi N}^{(\slashed P T)} & = - C_{VA}^{(v)} F_A^{(v)} \, , \\
        C_{8, \chi N}^{(\slashed P T)} & = C_{AV}^{(s)} F_1^{(s)} \, , \\
        C_{9, \chi N}^{(\slashed P T)} & = - C_{VA}^{(s)} F_A^{(s)} \, , \\
        C_{10, \chi N}^{(\slashed P T)} & = C_{AV}^{(v)} F_1^{(v)} \, . 
    \end{align}
Lastly, the LECs of the $\slashed P \slashed T$ operators are related to the quark level couplings according to
    \begin{align}
        C_{11, \chi N}^{(\slashed P \slashed T)} & = \frac{1}{2} C_{SP}^{(v)} F_P^{(v)} + \frac{1}{m_N^2 m_\chi} C_{TT}^{(v)'} F_{T,2}^{(v)} \, , \\
        C_{12, \chi N}^{(\slashed P \slashed T)} & = -\frac{1}{2m_\chi} C_{PS}^{(s)} \sigma_{\pi N} + \frac{1}{2m_N} C_{TT}^{(s)'} \left( F_{T,0}^{(s)} + F_{T,1}^{(s)} \right)  \, , \\
        C_{13, \chi N}^{(\slashed P \slashed T)} & = -\frac{1}{2m_\chi} C_{PS}^{(v)} \delta m_N + \frac{1}{2m_N} C_{TT}^{(v)'} \left( F_{T,0}^{(v)} + F_{T,1}^{(v)} \right) \, , \\
        C_{14, \chi N}^{(\slashed P \slashed T)} & = \frac{1}{2m_N} C_{SP}^{(s)} F_P^{(s)} + \frac{1}{m_N^2 m_\chi} C_{TT}^{(s)'} F_{T,2}^{(s)} \, , \\
        C_{15, \chi N}^{(\slashed P \slashed T)} & = -C_{TT}^{(v)'} F_{T,0}^{(v)} + \frac{1}{m_N^2 m_\chi} C_{TT}^{(v)'} F_{T,2}^{(v)} \, , \\
        C_{16, \chi N}^{(\slashed P \slashed T)} & = -C_{TT}^{(s)'} F_{T,0}^{(s)} + \frac{1}{m_N^2 m_\chi} C_{TT}^{(s)'} F_{T,2}^{(s)} \, .
    \end{align}
Finally, we reiterate a point from Ref.~\cite{bishara_from-quarks_2017}: it is inconsistent to turn off all but one coupling in the nucleon level EFT. 
For example, the operators proportional to $C_{5, 
\chi N}^{(\slashed P T)}$ and $C_{7, 
\chi N}^{(\slashed P T)}$ are generated by the same operator at the quark level.
Therefore, if one of these operators is included in an analysis the other one is also required.

%% file: Appendix/matching2.tex
\section{Two-body currents with dibaryons}
    \label{appendix:dibaryon}

The couplings of two-body currents to the external dark matter current in \eftnopi with dibaryon fields can be matched to the theory without dibaryon fields by matching the amplitudes for dark matter-dibaryon scattering.
The interactions between the external WIMP and the dibaryon fields appear in Eq.~\eqref{eq:triton:dibaryon_lagrangian}
\begin{equation}
        \tag{\ref{eq:triton:dibaryon_lagrangian}}
\begin{aligned}
    \calL_\text{dibaryon} & = l^\chi_1  \chi^\dagger \chi  t_i^\dagger t_i + il^\chi_2\epsilon_{ijk} \left( \chi^\dagger \sigma^i \chi \right) t_j^\dagger t_k + l^\chi_3 \left( \chi^\dagger \sigma^i \chi \right) \left( s_3^\dagger t_i + \text{H.c.} \right) + l^\chi_4 \left( \chi^\dagger \chi \right) s_a^\dagger s_a \\
    & + l^\chi_5 i \epsilon^{3ab}  \left( \chi^\dagger \chi \right) s_a^\dagger s_b + l^\chi_6 \mathcal{I}_{ab}\left( \chi^\dagger \chi \right) s_a^\dagger s_b + l^\chi_7 \left( \chi^\dagger \sigma^i \chi \right) \left( i s_3^\dagger t_i + \text{H.c.} \right) .
    \end{aligned}
\end{equation}
To be concrete, consider WIMP-deuteron elastic scattering again but with dibaryon fields.
A NLO calculation in the Z parameterization leads to
    \begin{align}
        i \calM & = -Z \frac{8 \gamma}{q} \tan^{-1} \left( \frac{q}{4 \gamma} \right) \left( i \delta_{ij} \delta_{rs} C_{1, \chi N}^{(PT)} + \epsilon_{ijk} \sigma^k_{rs} C_{3, \chi N}^{(PT)} \right)  - i l^\chi_1 \delta_{ij} \delta_{rs} \frac{8 \pi \gamma}{m^2 y^2} + l_2^\chi \epsilon_{ijk} \sigma^k_{rs} \frac{8 \pi \gamma}{m^2 y^2} \, .
    \end{align}
The LECs $l_1^\chi$ and $l_2^\chi$ can be determined in terms of the LEC combinations $C_{1, \chi \NN}^{(\text{SI}, s)} + C_{2 \chi \NN}^{(\text{SI}, s)}$ and $C_{1, \chi \NN}^{(\text{SD}, s)}$ by matching the amplitudes in $q^2 \to 0$ limit. 
This leads to the relations
    \begin{align}
        l_1^\chi & = \frac{m_N^2 y_t^2}{4 \pi \gamma} C_{1, \chi N}^{(PT)} \left( Z - 1 \right) + \frac{m_N^2 y_t^2}{8 \pi^2} \left( \mu - \gamma \right)^2 (C_{1, \chi \NN}^{(\text{SI}, s)} + C_{2 \chi \NN}^{(\text{SI}, s)}) \, , \label{eq:dibaryon:l1} \\
        l_2^\chi & = \frac{m_N^2 y_t^2}{4 \pi \gamma} C_{3, \chi N}^{(PT)} \left( Z - 1 \right) + \frac{m_N^2 y_t^2}{8 \pi^2} \left( \mu - \gamma \right)^2 C_{1, \chi \NN}^{(\text{SD}, s)} \label{eq:dibaryon:l2} \, .
    \end{align}
In both matching relations, the second term scales in the way with \Nc as the first term.
Calculating analogous amplitudes shows that this is a general feature, although knowing the exact form of the relationships is not imperative at this stage. Therefore, we may apply the large-\Nc scalings of the two-nucleon currents from Eq.~\eqref{eq:Lagrangian:two_nucleon_partial} to the corresponding LECs in the dibaryon formalism, which indicates the scalings
    \begin{align}
        l^\chi_3 & \propto C_{1, \chi \NN}^{(\text{SD}, v)} \sim O(\Nc) \, , \label{eq:dibaryon:l3} \\
        l^\chi_4 & \propto C_{1, \chi \NN}^{(\text{SI}, s)} - 3 C_{2 \chi \NN}^{(\text{SI}, s)} \sim O(\Nc) \, , \label{eq:dibaryon:l4} \\
        l^\chi_5 & \propto C_{1, \chi \NN}^{(\text{SI}, v)} \sim O(1) \, , \label{eq:dibaryon:l5} \\
        l^\chi_6 & \propto C_{1, \chi \NN}^{(\text{SI}, t)} \sim O(\Nc) \, , \label{eq:dibaryon:l6} \\
        l^\chi_7 & \propto C_{2, \chi \NN}^{(\text{SD}, v)} \sim O(1) \, . \label{eq:dibaryon:l7}
    \end{align}
Furthermore, the LECs in the dibaryon formulation are renormalization group invariant.

From the definition of $x$ in Eq.~\eqref{eq:cross_sections:deuteron:x}, we also find
    \begin{align}
        l_1^\chi & = \frac{m_N}{\gamma} C^{(s)}_{SS} \sigma_{\pi N} (Z-1) \left( \frac{x}{2} + 1 \right) \, .
    \end{align}
Thus, setting $l_1 \approx 14.67 \sigma_{\pi N} C^{(s)}_{SS}$, we may use the large-\Nc scalings to estimate the values of the remaining LECs in order to predict the remaining response functions in the three-body system.
We expect $l^\chi_3$, $l^\chi_4$, and $l^\chi_6$ to be roughly the same size and $l^\chi_2$, $l^\chi_5$, $l^\chi_7$ to be about 1/3 of $l^\chi_1$ if a coupling is also turned on at the quark level that generates spin dependent scattering.